\documentclass{article}
\usepackage{romp}

\usepackage[dvips]{graphicx}
\usepackage{amsmath}
\usepackage{amsfonts}
\usepackage{amssymb}
\usepackage{longtable} 
\usepackage{color}

\allowdisplaybreaks[4] 

 \renewcommand{\d}{{\rm d}} 

\newcommand{\df}{\ {\overset {\rm def} =}\ }
\newcommand{\dr}[2]{\frac {{\rm d} {#1}} {{\rm d} {#2}}}

\newcommand{\dril}[2]{{{\rm d} {#1}} / {{\rm d} {#2}}}
\newcommand{\pdril}[2]{{\partial {#1}} / {\partial {#2}}}

\newcommand{\llim}[1] {\ {\underset {#1} {\longrightarrow}}\ }

\title{Expansion of bundles of light rays \\
in the Lema\^{\i}tre -- Tolman models}
\author{Andrzej Krasi\'nski \\ N. Copernicus Astronomical Centre, Polish Academy
of Sciences \\
Bartycka 18, 00 716 Warszawa, Poland \\
e-mail: akr@camk.edu.pl}

\begin{document}

\maketitle

\begin{abstract}
The locus of $\theta \df {k^{\mu}};_{\mu} = 0$ for bundles of light rays emitted
at noncentral points is investigated for Lema\^{\i}tre -- Tolman (L--T) models.
The three loci that coincide for a central emission point: (1) maxima of $R$
along the rays, (2) $\theta = 0$, (3) $R = 2M$ are all different for a
noncentral emitter. If an extremum of $R$ along a nonradial ray exists, then it
must lie in the region $R > 2M$. In $2M < R \leq 3M$ it can only be a maximum;
in $R > 3M$ both minima and maxima can exist. The intersection of (1) with the
equatorial hypersurface (EHS) $\vartheta = \pi/2$ is numerically determined for
an exemplary toy model (ETM), for two typical emitter locations. The equation of
(2) is derived for a general L--T model, and its intersection with the EHS in
the ETM is numerically determined for the same two emitter locations. Typically,
$\theta$ has no zeros or two zeros along a ray, and becomes $+ \infty$ at the
Big Crunch (BC). The only rays on which $\theta \to - \infty$ at the BC are the
radial ones. Along rays on the boundaries between the no-zeros and the two-zeros
regions $\theta$ has one zero, but still tends to $+ \infty$ at the BC. When the
emitter is sufficiently close to the center, $\theta$ has 4 or 6 zeros along
some rays (resp. 3 or 5 on the boundary rays). For noncentral emitters in a
collapsing L--T model, $R = 2M$ is still the ultimate barrier behind which
events become invisible from outside; loci (1) and (2) are not such barriers.
\end{abstract}

\noindent {\bf Keywords:} general relativity, cosmological models, light
propagation, horizons.

\section{Motivation and background}\label{intro}

\setcounter{equation}{0}

We are interested in the outer boundary of a set whose every point lies in a
trapped surface (the latter is a 2-surface whose family of outgoing orthogonal
light rays has nonpositive expansion scalar). With a slight abuse of the
original definition \cite{HaEl1973} we will refer to this as the apparent
horizon (AH).

In the Lema\^{\i}tre \cite{Lema1933} -- Tolman \cite{Tolm1934} (L--T) models the
AH has been so far considered only for bundles of light rays emitted at the
world line of the central observer \cite{KrHe2004b,PlKr2006}. In this case, the
AH can be defined in two ways:

(1) As the locus where the surface areas of the light fronts of the bundles
achieve maxima. At the same locus the areal radius $R$ of the light front
becomes maximum.

(2) As the locus where the expansion scalars $\theta \df {k^{\mu}};_{\mu}$ of
such bundles become zero ($k^{\mu}$ is the vector field tangent to the rays).

\noindent Both these definitions determine the same hypersurface $R = 2M$.

\begin{figure}[h]
 \hspace{-3mm} \includegraphics[scale = 0.65]{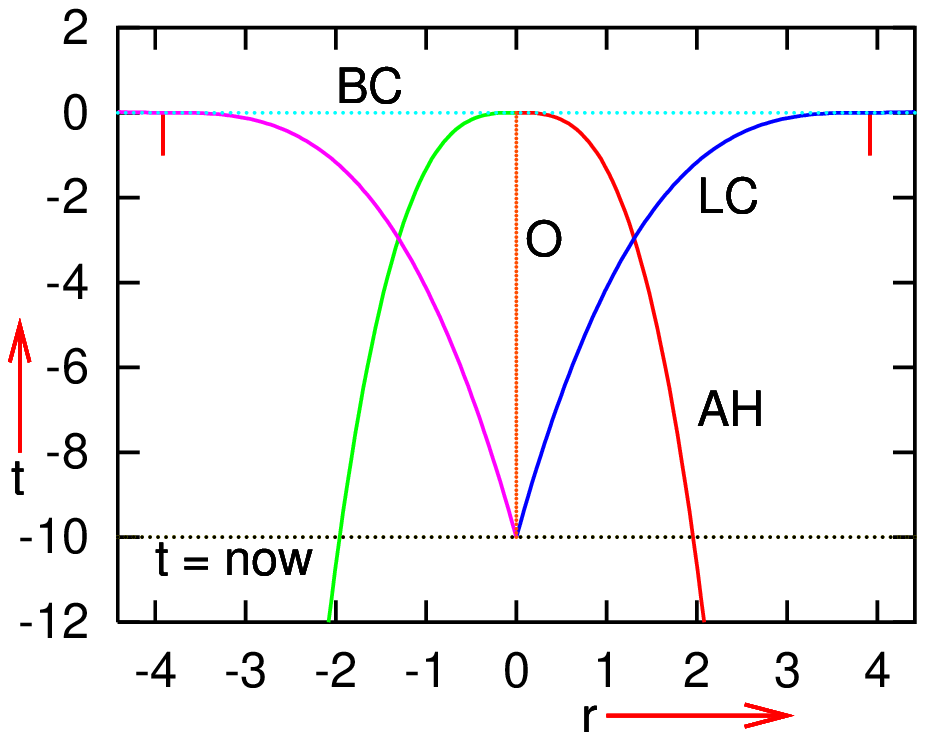}
\includegraphics[scale = 0.65]{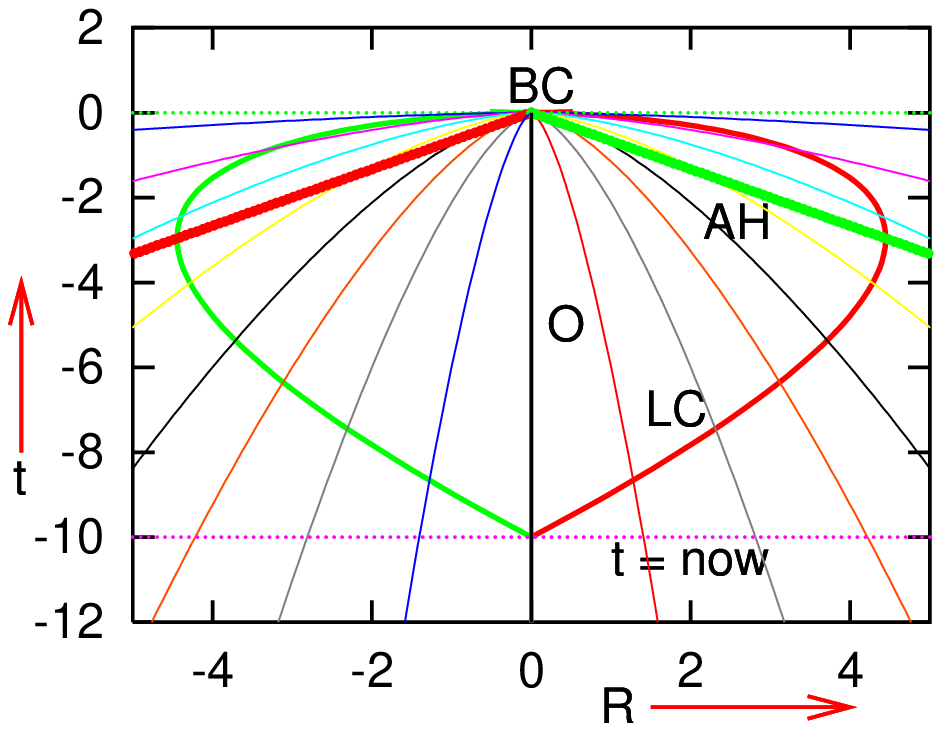}
\caption{{\bf Left panel:} Profiles of the future apparent horizon AH of a
comoving observer O and of the future light cone LC of O's present instant in
the collapsing Friedmann model (\ref{1.1}). {\bf Right panel:} The same
situation in the $(t, R)$ variables where $R = r S(t)$. At the AH $R$ becomes
maximum along the rays. See the text for more explanation.}
 \label{AHflat}
\end{figure}

This created the impression that the AH so defined is common to all light
emitters. However, in Friedmann models \cite{Frie1922}, which are the spatially
homogeneous limits of L--T models, each observer is central because of the
homogeneity and each one has a differently located AH. The exemplary model used
for Fig. \ref{AHflat} has the metric
\begin{equation}\label{1.1}
{\rm d} s^2 = {\rm d} t^2 - S^2(t) \left[{\rm d} r^2 + r^2 \left({\rm d}
\vartheta^2 + \sin^2 \vartheta {\rm d} \varphi^2\right)\right]
\end{equation}
with $S(t) \propto (t_{\rm BC} - t)^{2/3}$, $t_{\rm BC}$ is the Big Crunch time.
In the left panel the $(t, r)$ coordinates are comoving. The future light cone
LC of the present instant of observer $O$ hits the BC tangentially at $r$ values
marked by the vertical strokes. In the right panel the coordinates are $t$ and
the areal radius $R = r S(t)$, the BC is a single point and the AH profile is
the pair of straight lines. The curves converging at the BC are world lines of
particles of the cosmic medium -- in the left panel they would be vertical
straight lines. Such structures exist around every comoving observer world line
in any collapsing Friedmann model.

So, it was puzzling where the AH would be for a noncentral observer in an L--T
model. The present paper aims at answering this question. The loci of extrema of
$R$ and of $\theta = 0$ in L--T models are here investigated for bundles of rays
that originate at noncentral events. Then, sets (1), (2) and (3) the locus of $R
= 2M$ are all different.\footnote{\label{noconfu} The loci of $\theta = 0$ and
of extremum of $D_a$ -- the area distance from the origin O of the ray bundle --
do coincide \cite{Perl2004}. But if O is not at the center of symmetry, then
$D_a \neq R$ and their extrema split.}. This is illustrated using the explicit
L--T toy-model (ETM), first introduced in Ref. \cite{KrHe2004b}.

In Sec. \ref{bapro}, basic information about the L--T models is given. In Sec.
\ref{LTnullgeo}, the equations of null geodesics in these models are written out
and prepared to numerical integration. In Sec. \ref{Rextre}, the equation
defining a local extremum of the areal radius $R$ along a light ray is discussed
for a general L--T model. It is shown that on nonradial rays an extremum can
exist only in the $R > 2M$ region. If it occurs in $R < 3M$, then it is
necessarily a maximum. In $R > 3M$ both minima and maxima are possible (but need
not exist).

In Sec. \ref{Rextreexam} the loci of extrema of $R$ are discussed for the ETM.
They are numerically calculated for rays running in the equatorial hypersurface
$\vartheta = \pi/2$ (EHS), in the recollapse phase of the model. On some
nonradial rays $R$ monotonically decreases to 0 achieved at the BC. On some
other rays, $R$ has only maxima, on still other ones it has both minima and
maxima. The latter can happen when the ray leaves the light source toward
decreasing $R$ (which is impossible when the source is at the center where $R =
0$).

In Sec. \ref{extexp}, the equation of the locus of $\theta = 0$ for a bundle of
light rays in a general L--T model is derived. Except on outward radial rays, it
does not coincide with the locus of an extremum of $R$. A method to numerically
calculate $\theta$ along a nonradial ray is given; an auxiliary nearby ray is
needed for that.

In Sec. \ref{exemzerotheta}, the $\theta = 0$ equation is numerically solved for
rays running in the EHS of the ETM used in Sec. \ref{Rextreexam}. Typically,
$\theta$ has no zero or two zeros along a ray, and becomes $+ \infty$ at the Big
Crunch (BC), so the ray bundle is infinitely {\em diverging} at the BC. The only
rays on which $\theta \to - \infty$ at the BC are the radial ones. Along rays on
the boundaries between the no-zeros and the two-zeros regions $\theta$ has one
zero, but still tends to $+ \infty$ at the BC. When the emitter is sufficiently
close to the center, $\theta$ has 4 or 6 zeros along rays passing near the
center (resp. 3 or 5 on the boundary rays). The loci of $\theta$-zeros and
temporal orderings of loci (1) -- (3) along various rays are displayed for
exemplary emission points of the ray bundles. A locus of $\theta = 0$ may lie
earlier or later than $R = 2M$ and than the maximum of $R$, depending on the
initial direction of the ray.

In Sec. \ref{sumcon} the results of the paper are summarised and discussed. One
of the conclusions is that the hypersurface $R = 2M$ still is an AH for
noncentral emitters. Namely, if $\theta = 0$ occurs at a point $p_1$ in the
region $R > 2M$, then the radial ray sent outwards from $p_1$ will proceed some
distance toward larger $R$ -- which means that $p_1$ is not yet locally trapped.
On the other hand, if $\theta = 0$ occurs at $p_2$ in the region $R < 2M$, then
events along this ray became invisible from outside before the ray reached
$p_2$. Thus, in a collapsing L--T model, $R = 2M$ is the ultimate barrier from
behind which no light rays can get to the outside world; the loci of maximum $R$
and of $\theta = 0$ are not such barriers.

 \section{Basic properties of the Lema\^{\i}tre-Tolman models}\label{bapro}

\setcounter{equation}{0}

The L-T models \cite{Lema1933,Tolm1934,PlKr2006} are spherically symmetric
nonstatic solutions of the Einstein equations with a dust source. Their metric
is
 \begin{equation}\label{2.1}
\d s^2 = \d t^2 - \frac {{R,_r}^2}{1 + 2E(r)} \d r^2 - R^2(t,r)(\d\vartheta^2 +
\sin^2\vartheta\d\varphi^2),
 \end{equation}
where $R(t,r)$ is determined by the equation
 \begin{equation}\label{2.2}
{R,_t}^2 = 2E(r) + 2M(r) / R + \Lambda R^2 / 3;
 \end{equation}
$E(r) \geq -1/2$ and $M(r)$ are arbitrary functions, $R,_r \df \pdril R r$,
$R,_t \df \pdril R t$, and $\Lambda$ is the cosmological constant. The
mass-density is
 \begin{equation}\label{2.3}
 \kappa \rho = \frac {2M,_r} {R^2 R,_r}, \qquad
 \text{where}\ \kappa \df \frac {8\pi G} {c^4}.
 \end{equation}
In the following we assume $\Lambda = 0$. Then (\ref{2.2}) can be solved in
terms of elementary functions \cite{PlKr2006}. We will use only an $E < 0$
solution, in which
 \begin{eqnarray}
R(t,r) &=& \displaystyle{\frac{M}{(-2E)}} (1 - \cos\eta), \label{2.4} \\
\eta - \sin\eta &=& \displaystyle{\frac {(-2E)^{3/2}}{M}} \left[t -
t_B(r)\right], \label{2.5}
 \end{eqnarray}
where $\eta$ is a parameter and the arbitrary function $t_B(r)$ determines the
local Big Bang (BB) instant at $t = t_B(r)$. This model is initially expanding
and later collapses to the final singularity (BC) at $t = t_C(r)$. Writing
(\ref{2.5}) at $\eta = 2\pi$ where $t = t_C$ we obtain
\begin{equation}\label{2.6}
2 \pi = \displaystyle{\frac {(-2E)^{3/2}}{M}} \left[t_C(r) - t_B(r)\right],
\end{equation}
so of the four functions $M$, $E$, $t_B$ and $t_C$ only three are independent.

All the formulae above are covariant under the transformations $\tilde{r} =
g(r)$, so we can give one of the three functions $E(r)$, $M(r)$ and ($t_B(r)$ or
$t_C(r)$) a convenient shape. In our exemplary toy model introduced in Sec.
\ref{Rextre}, it will be convenient to take $\tilde{r} = M(r)$.\footnote{Such a
choice of the radial coordinate is allowed in those ranges of $r$ where the
function $M(r)$ is monotonic. In the model of Sec. \ref{Rextre} this problem
does not arise as the range of $M$ is $[0,\infty)$ and no other radial
coordinate appears. \label{glupiref}}

The Friedmann models \cite{Frie1922} are contained in the L--T class as the
limit:
\begin{equation}\label{2.7}
t_B = \text{constant}, \qquad |E|^{3/2}/M = \text{constant},
\end{equation}
and with $M = {\rm constant} \times r^3$ their most popular coordinate
representation results.

Shell crossings (SCs), are loci at which neighbouring constant-$r$ shells
collide. At a SC, $R,_r = 0 \neq 1 + 2E$; they are curvature singularities. The
conditions on $M$, $E$ and $t_B$ that ensure the absence of SCs were given in
Ref. \cite{HeLa1985}, and will be used below.

\section{Light rays in an L--T model}\label{LTnullgeo}

\setcounter{equation}{0}

The tangent vectors $k^{\alpha} = \dril {x^{\alpha}} {\lambda}$ to geodesics of
the metric (\ref{2.1}) obey
\begin{eqnarray}
&& \dr {k^t} {\lambda} + \frac {R,_r R,_{t r}} {1 + 2E}\ \left(k^r\right)^2 + R
R,_t\ \left[\left(k^{\vartheta}\right)^2 + \sin^2\vartheta\
\left(k^{\varphi}\right)^2\right] = 0, \label{3.1} \\
&& \dr {k^r} {\lambda} + 2 \frac {R,_{t r}} {R,_r}\ k^t k^r + \left(\frac {R,_{r
r}} {R,_r} - \frac {E,_r} {1 + 2E}\right)\ \left(k^r\right)^2 \nonumber \\
&& \ \ \ \  \ \ \ \  \ \ \ \ - \frac {(1 + 2E) R} {R,_r}\
\left[\left(k^{\vartheta}\right)^2 + \sin^2\vartheta\
\left(k^{\varphi}\right)^2\right] = 0, \label{3.2} \\
&& \dr {k^{\vartheta}} {\lambda} + 2 \frac {R,_t} R\ k^t k^{\vartheta} + 2 \frac
{R,_r} R\ k^r k^{\vartheta} - \cos \vartheta\sin \vartheta
\left(k^{\varphi}\right)^2 = 0, \label{3.3} \\
&& \dr {k^{\varphi}} {\lambda} + 2 \frac {R,_t} R\ k^t k^{\varphi} + 2 \frac
{R,_r} R\ k^r k^{\varphi} + 2 \frac {\cos \vartheta} {\sin \vartheta}\
k^{\vartheta} k^{\varphi} = 0, \label{3.4}
\end{eqnarray}
where $\lambda$ is the affine parameter. The geodesics are null when
\begin{equation}\label{3.5}
\left(k^t\right)^2 - \frac {{R,_r}^2 \left(k^r\right)^2} {1 + 2E} - R^2
\left[\left(k^{\vartheta}\right)^2 + \sin^2 \vartheta
\left(k^{\varphi}\right)^2\right] = 0.
\end{equation}
Since $R,_t k^t + R,_r k^r = \dril R {\lambda}$, the general solution of
(\ref{3.4}) is
\begin{equation}\label{3.6}
R^2 \sin^2 \vartheta k^{\varphi} = J_0,
\end{equation}
where $J_0$ is constant along the geodesic. The case $J_0 = 0$ corresponds to
two situations:

(a) $\vartheta = 0, \pi$; then the ray stays on the axis of symmetry, with
undetermined $\varphi$.

(b) $k^{\varphi} = 0$ with $\vartheta$ (as yet) unspecified; then the ray stays
in a constant-$\varphi$ hypersurface.

Using (\ref{3.6}), the general solution of (\ref{3.3}) is
\begin{equation}\label{3.7}
R^4 \left(k^{\vartheta}\right)^2 \sin^2 \vartheta + {J_0}^2 = C^2 \sin^2
\vartheta,
\end{equation}
where $C$ is another constant along the geodesic. When $C = 0$, the geodesic is
radial. Then $J_0 = 0$ and either (a) $\vartheta = 0, \pi$, or (b) $\vartheta$
is constant along the ray and $\varphi$ is constant by (\ref{3.6}). When $C =
\pm J_0 \neq 0$, the geodesic remains in the equatorial hypersurface $\vartheta
= \pi/2$ (which is not flat even when $E = 0$). For later reference let us note
the following:
\begin{equation}\label{3.8}
 \parbox{12cm} {
{\bf The coordinates $(\vartheta, \varphi)$ can be adapted to any single
geodesic so that it stays in the hypersurface $\vartheta' = \pi / 2$ in the new
coordinates $(\vartheta', \varphi')$.}} \
\end{equation}
This is a consequence of spherical symmetry of the spacetime \cite{PlKr2006}.
Equation (\ref{3.7}) implies
\begin{equation}\label{3.9}
C^2 \sin^2 \vartheta \geq {J_0}^2.
\end{equation}

For rays with $J_0 \neq 0$, eq. (\ref{3.6}) implies in addition:
\begin{equation}\label{3.10}
k^{\varphi} \equiv \dr {\varphi} {\lambda} \to \infty \qquad {\rm when} \qquad R
\to 0.
\end{equation}
Thus, if $|\dril r {\lambda}| < \infty$ at the intersection with the BB or BC,
then $\dril {\varphi} r \llim{t \to t_B} \infty$, i.e. these rays meet the
singularity being tangent to a surface of constant $r$.

Equations (\ref{3.3}) and (\ref{3.4}) are now solved. From (\ref{3.6}) and
(\ref{3.7}) we get
\begin{equation}\label{3.11}
\left(k^{\vartheta}\right)^2 + \sin^2 \vartheta \left(k^{\varphi}\right)^2 = C^2
/ R^4,
\end{equation}
and then (\ref{3.5}) becomes
\begin{equation}\label{3.12}
\left(k^t\right)^2 = \frac {{R,_r}^2 \left(k^r\right)^2} {1 + 2E} + \frac {C^2}
{R^2}.
\end{equation}
Equations (\ref{3.1}) -- (\ref{3.2}), using (\ref{3.6}) -- (\ref{3.12}),
simplify to
\begin{eqnarray}
\dril t {\lambda} &=& k^t, \label{3.13} \\
\dr {k^t} {\lambda} &=& \left[(C / R)^2 - \left(k^t\right)^2\right]
\frac {R,_{t r}} {R,_r} - \frac {C^2 R,_t} {R^3}, \label{3.14} \\
\dril r {\lambda} &=& k^r, \label{3.15} \\
\dr {k^r} {\lambda} &=& - 2 \frac {R,_{t r}} {R,_r}\ k^t k^r - \left(\frac
{R,_{rr}} {R,_r} - \frac {E,_r} {1 + 2E}\right)\ \left(k^r\right)^2 + \frac {C^2
(1 + 2E)} {R^3 R,_r}. \label{3.16}
\end{eqnarray}
The initial data for (\ref{3.13}) -- (\ref{3.16}) are $t, r, k^t$ and $k^r$ at
the initial point of the ray $(t, r, \vartheta, \varphi) = (t_o, r_o,
\vartheta_o, \varphi_o)$. In numerical calculations, (\ref{3.12}) will be used
at every step to correct the value of $k^t$ found by integrating (\ref{3.13}) --
(\ref{3.16}).

One more initial condition is achieved by rescaling $\lambda$:
\begin{equation}
k^t(t_o) = \pm 1 \label{3.17}
\end{equation}
($+$ for future-directed, $-$ for past-directed rays). With (\ref{3.17}) we have
from (\ref{3.12}),
\begin{equation}\label{3.18}
C^2 \leq R^2(t_o,r_o) \df {R_o}^2;
\end{equation}
the equality occurs when $k^r(r_o) = 0$.

The following formulae \cite{PlKr2006} are useful in numerical calculations:
\begin{eqnarray}
R,_r &=& \left(\frac {M,_r} M - \frac {E,_r} E\right)R + \left[\left(\frac 3 2
\frac {E,_r} E - \frac {M,_r} M\right) \left(t - t_B\right) - t_{B,r}\right]
R,_t, \label{3.19} \\
R,_{tr} &=& \frac {E,_r} {2E}\ R,_t - \frac M {R^2}\ \left[\left(\frac 3 2 \frac
{E,_r} E - \frac {M,_r} M\right) \left(t - t_B\right) - t_{B,r}\right],
\label{3.20} \\
R,_{rr} &=& \left[\frac {M,_{rr}} M - \frac {2 M,_r E,_r} {ME} + \frac
{2 {E,_r}^2} {E^2} - \frac {E,_{rr}} E\right] R \nonumber \\
&+& \left[\left(\frac 3 2 \frac {E,_{rr}} E - \frac {9 {E,_r}^2} {4 E^2} + \frac
{2 M,_r E,_r} {ME} - \frac {M,_{rr}} M\right) \left(t -
t_B\right) - \frac {E,_r} E\  t_{B,r} - t_{B,rr}\right] R,_t, \nonumber \\
&-& \frac M {R^2}\ \left[\left(\frac 3 2 \frac {E,_r} E - \frac {M,_r} M\right)
\left(t - t_B\right) - t_{B,r}\right]^2. \label{3.21}
\end{eqnarray}

\section{The extremum of $R$ along a ray}\label{Rextre}

\setcounter{equation}{0}

The following holds at an extremum of $R(t, r)$ along the curve tangent to
$k^{\alpha}$
\begin{equation}\label{4.1}
\dr R {\lambda} = R,_t k^t + R,_r k^r = 0.
\end{equation}
On future-directed curves $k^t > 0$, so $R,_t k^t > 0$ when the model expands
and $R,_t k^t < 0$ when it collapses. When shell crossings and necks are absent,
$R,_r > 0$ \cite{HeLa1985}. Thus, solutions of (\ref{4.1}) may exist only where
$R,_t k^r < 0$. On a null geodesic, (\ref{4.1}) implies, via (\ref{3.12})
\begin{equation}\label{4.2}
{R,_t}^2 \left[\frac {\left(R,_r k^r\right)^2} {1 + 2E} + \frac {C^2}
{R^2}\right] = \left(R,_r k^r\right)^2.
\end{equation}
Using (\ref{2.2}) with $\Lambda = 0$ in this we obtain
\begin{equation}\label{4.3}
\left(\frac {2M} R - 1\right) \left(R,_r k^r\right)^2 + \left(\frac {CR,_t}
R\right)^2\ (1 + 2E) = 0.
\end{equation}
Where $R < 2M$, both terms in (\ref{4.3}) are non-negative, so (\ref{4.3}) may
hold only when both are zero. The only physically meaningful situation when this
happens is
\begin{equation}\label{4.4}
R,_r = 1 + 2E = 0.
\end{equation}
This is a neck \cite{PlKr2006}. Apart from this locus, $1 + 2E > 0$ must hold in
order that the signature is the physical $(+, -, -, -)$. Other solutions of
(\ref{4.3}) do not exist with $R < 2M$ because

(1) When $C = 0 = k^r$, the geodesic is timelike, while the locus of $C = R,_r =
0 < 1 + 2E$ is a shell crossing which we assumed not to exist.

(2) $R,_t = 0$ holds at maximum expansion (when $E < 0$), but then $R \geq 2M$
\cite{PlKr2006}.

\noindent Thus, the solution of (\ref{4.3}) is $R = 2M$ only on radial
geodesics, where $C = 0$. With $C \neq 0$, (\ref{4.3}) can have solutions only
where $R > 2M$ (see examples in Sec. \ref{Rextreexam}).

Given the value of $C$ and an initial point $p_0$, Eqs. (\ref{3.12}) --
(\ref{3.16}) with (\ref{3.6}) -- (\ref{3.7}) define a single ray, and then each
solution of (\ref{4.1})\footnote{On a given ray, (\ref{4.1}) may have more than
one solution or no solutions. See examples further on.} defines a point on that
ray. When the values of $C$ are changed with $p_0$ fixed, those points draw a
2-surface $S_{p_0}$. When $p_0$ is moved along an observer's world line, the
$S_{p_0}$ surfaces form a hypersurface $H$ in spacetime which touches $R = 2M$
along radial rays, but elsewhere lies in $R > 2M$. We will see in Sec.
\ref{extexp} that for a noncentral observer the hypersurface $H$ and the locus
of ${k^{\mu}};_{\mu} = 0$ do not coincide, see Secs. \ref{Rextreexam} and
\ref{exemzerotheta} for explicit examples.

Now we calculate $\dril {^2 R} {\lambda^2}$ from (\ref{4.1}). To eliminate
$\dril {k^t} {\lambda}$ and $\dril {k^r} {\lambda}$ we use (\ref{3.14}) and
(\ref{3.16}). We also use the derivative of (\ref{2.2}) with $\Lambda = 0$ by
$r$:
\begin{equation}\label{4.5}
R,_t R,_{tr} = \frac {M,_r} R - \frac {MR,_r} {R^2} + E,_r.
\end{equation}
The end result is
\begin{equation}\label{4.6}
\dr {^2 R} {\lambda^2} = - \frac {M,_r R,_r \left(k^r\right)^2} {R (1 + 2E)} +
\frac {C^2} {R^3}\left(1 - \frac {3M} R\right).
\end{equation}

For the model to be physical it is necessary that $\rho > 0$ in (\ref{2.3}), so
$M,_r R,_r > 0$ and the first term in (\ref{4.6}) is nonpositive. The second
term is negative (positive) where $R < 3M$ ($R > 3M$). Thus, $\dril {^2 R}
{\lambda^2} < 0$ where $R \leq 3M$, and if a solution of (\ref{4.1}) exists then
it is a maximum. Where $R > 3M$, the sign of $\dril {^2 R} {\lambda^2}$ depends
on the balance between the two terms in (\ref{4.6}), so both minima and maxima
of $R$ may exist;\footnote{For radial geodesics $C = 0$, so $\dril {^2 R}
{\lambda^2} < 0$ and only maxima are possible.} see Sec. \ref{Rextreexam}.

\section{Extrema of $R$ along nonradial rays in an exemplary L--T
model}\label{Rextreexam}

\setcounter{equation}{0}

\begin{figure}[h]
\includegraphics[scale = 0.6]{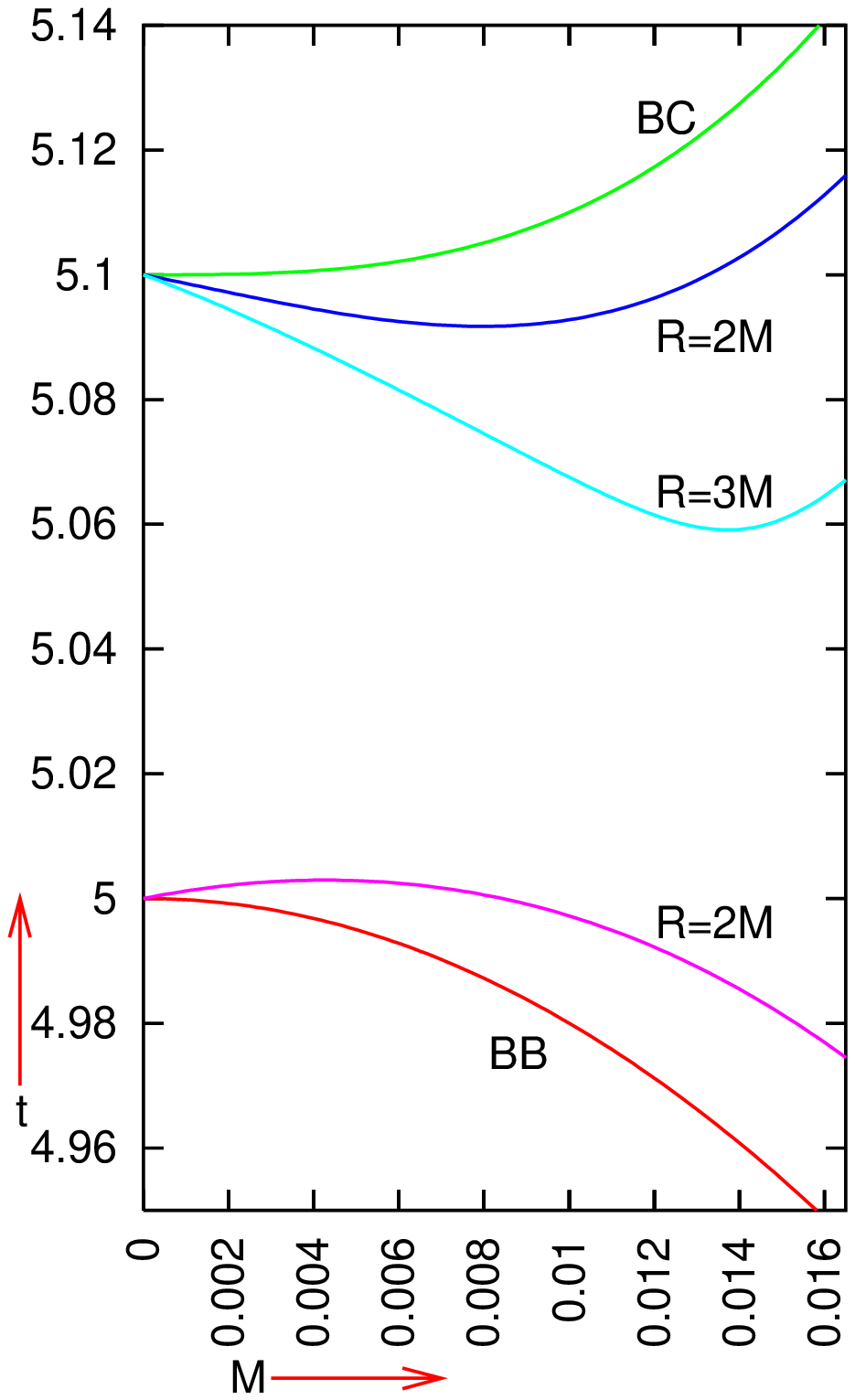}
\includegraphics[scale = 0.6]{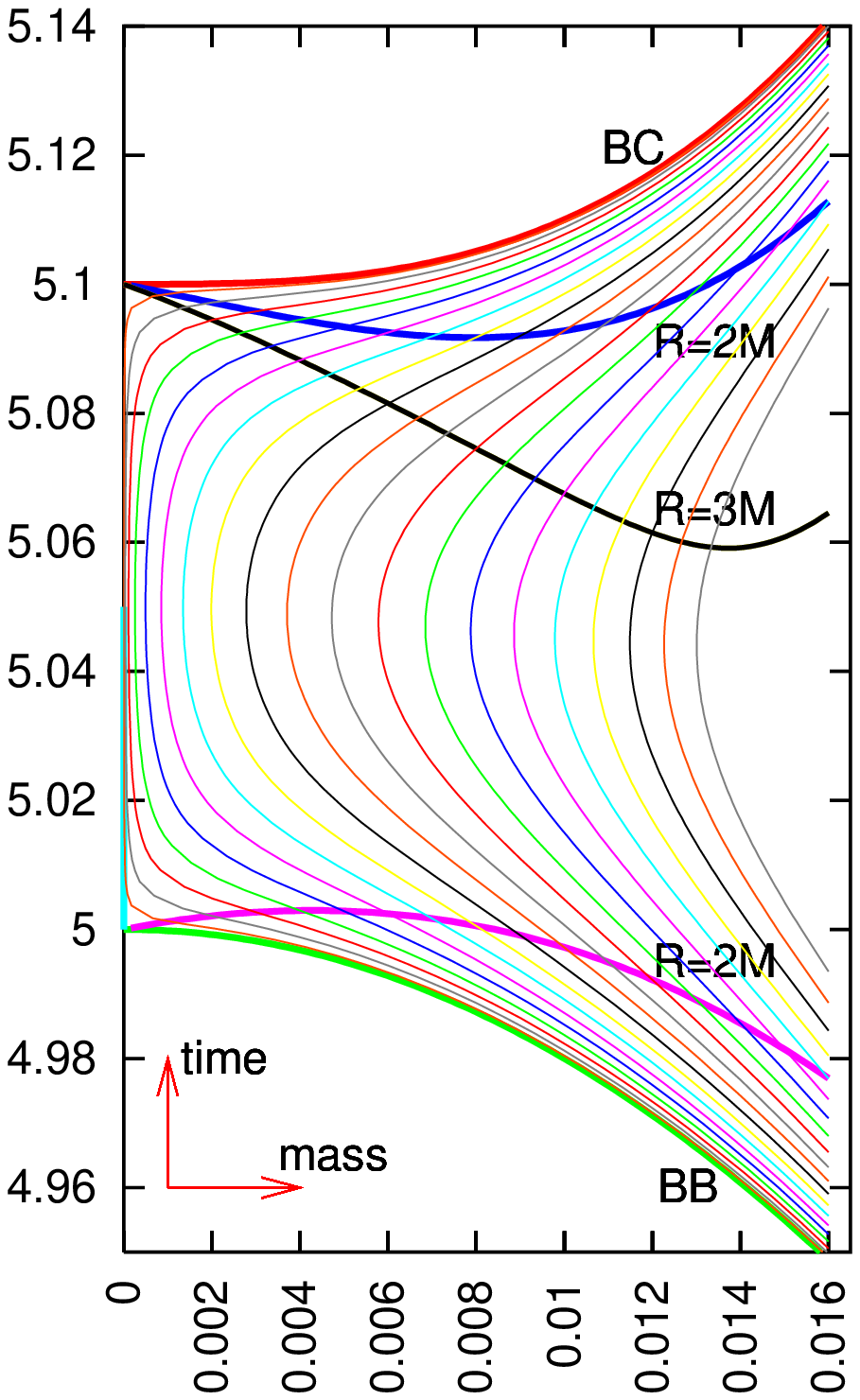}
\caption{{\bf Left panel:} The $t(M)$ profiles of the Big Bang (BB), Big Crunch
(BC), both $R = 2M$ sets and of the future $R = 3M$ set in the L--T model given
by (\ref{5.1}) -- (\ref{5.2}). {\bf Right panel:} Contours of constant $R$
written into the left panel. The $R = 0$ contour consists of the BB, the line $M
= 0$ and the BC. Values of $R$ increase from left to right at steps of 0.002,
from 0 to 0.04 on the rightmost contour.}
 \label{oldAH}
\end{figure}

To illustrate the conclusions of Sec. \ref{Rextre} we now consider a
recollapsing L--T toy model that has its Big Bang at $t = t_B(M)$ and its Big
Crunch at $t = t_C(M)$, where
\begin{eqnarray}
t_B(M) &=& - b M^2 + t_{B0}, \label{5.1} \\
t_C(M) &=& a M^3 + t_{B0} + T_0, \label{5.2}
\end{eqnarray}
with $a = 10^4$, $b = 200$, $t_{B0} = 5$ and $T_0 = 0.1$ being constants; the
mass function $M$ is used as the radial coordinate \cite{KrHe2004b,PlKr2006};
see footnote \ref{glupiref} in Sec. \ref{bapro}. This model is spatially
infinite and becomes spatially flat at $M \to \infty$; see Fig. \ref{oldAH}. Its
subspace $\vartheta = \pi/2$ can be imagined by rotating any panel of Fig.
\ref{oldAH} around the $M = 0$ axis. From (\ref{2.6}) we obtain

\begin{equation}\label{5.3}
E(M) = - \frac 1 2\ \left(\frac {2 \pi M} {t_C - t_B}\right)^{2/3} = - \frac 1
2\ \left(\frac {2 \pi M} {a M^3 + b M^2 + T_0}\right)^{2/3}.
\end{equation}

\begin{figure}[h]
 ${}$ \\[5mm]
\begin{center}
\hspace{-1cm} \includegraphics[scale = 0.7]{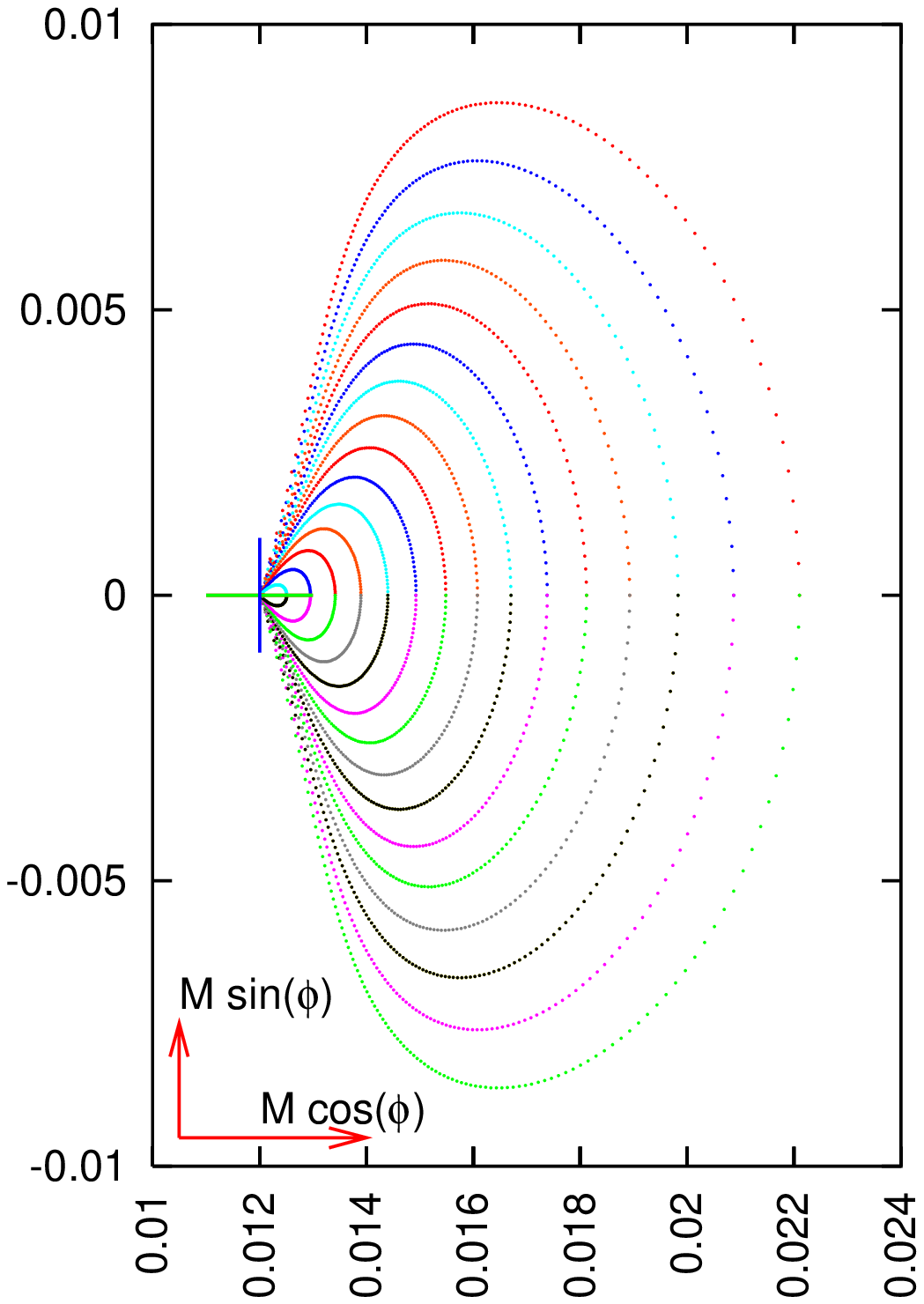}
 \hspace{2.5cm} \includegraphics[scale = 0.6]{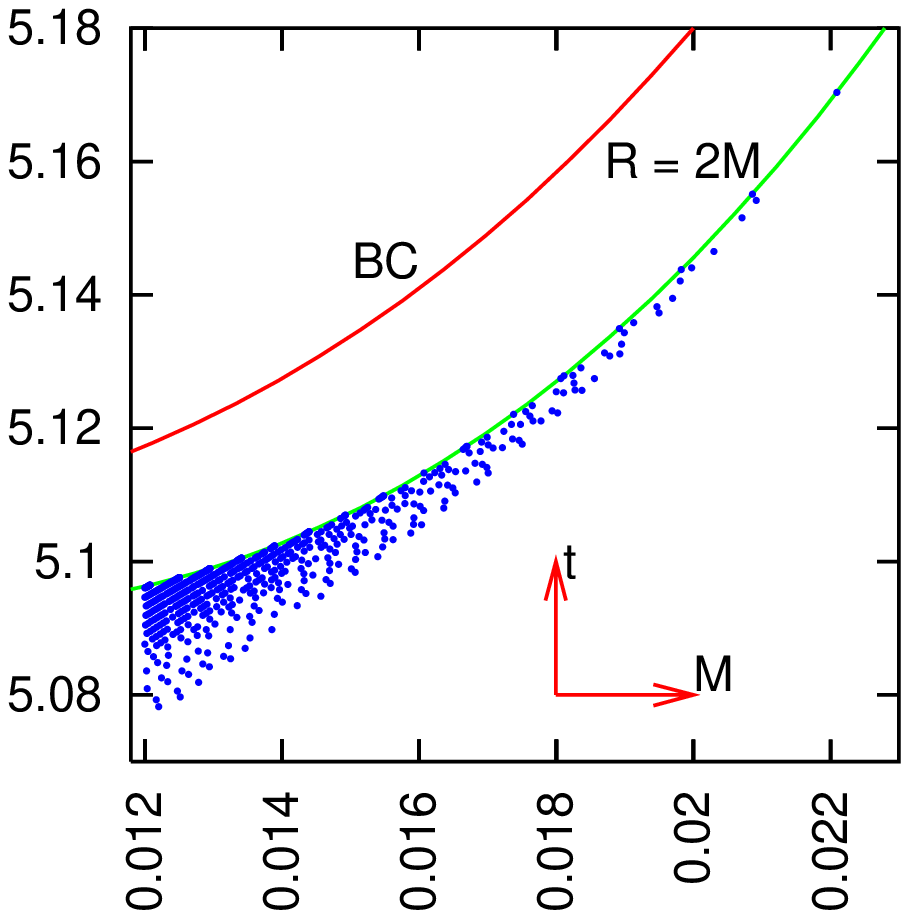}
\caption{{\bf Left:} Loci of $R$ maxima for 16 bundles of rays emitted from the
observer world line at $M = 0.012$ in the model of (\ref{5.1}) -- (\ref{5.2}),
projected on a $t =$ constant surface. The rays lie in the $\vartheta = \pi / 2$
hypersurface. The cross marks the $(M, \varphi) = (0.012, 0)$ coordinates of the
emitter. See the text for details. {\bf Right:} The $(M, t)$ coordinates of all
$R$ maxima.}
 \label{duzeringi}
\end{center}
\end{figure}

\begin{figure}[h]
 ${}$ \\[-2cm]
\begin{center}
 \hspace{5cm} \includegraphics[scale = 0.78]{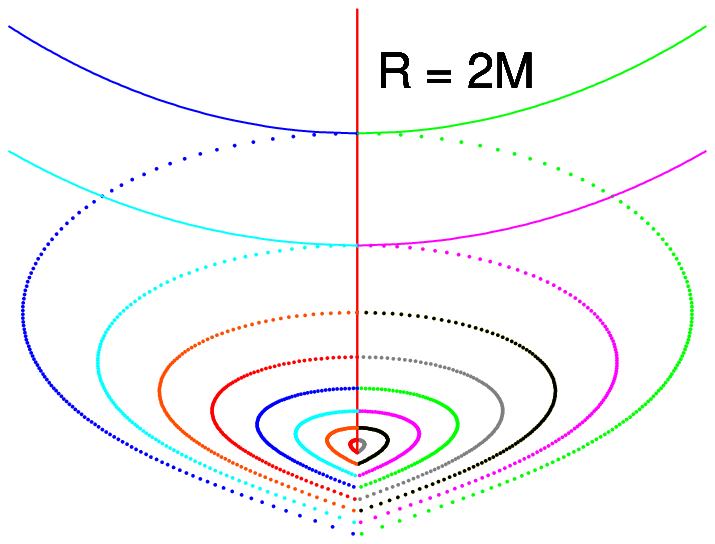}
 ${}$ \\[-10.4cm]
 \hspace{-7.5cm} \includegraphics[scale = 0.8]{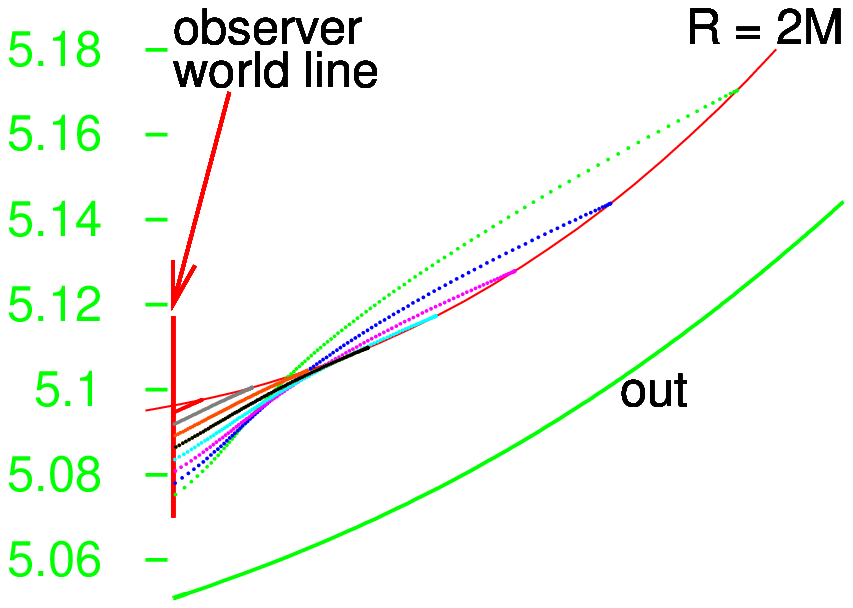}
 ${}$ \\[-2.3cm]
\caption{{\bf Left:} Projections of selected contours of $R$ maxima on the
surface $\varphi = 0$. The line marked ``out'' is a ray that escapes to $R =
\infty$ with no $R$ extrema. {\bf Right:} Projections of the same contours on
the surface $\varphi = \pi/2, 3 \pi / 2$. See the text for more comments. }
 \label{drawsecondproj}
\end{center}
\end{figure}

In this model let us consider the locus of extrema of $R$ along bundles of
future-directed rays emitted from the observer world line at $(M, \varphi) =
(0.012, 0)$. Figures \ref{duzeringi} and \ref{drawsecondproj} show this locus
for rays emitted at 16 points that run in the $\vartheta = \pi / 2$
hypersurface.\footnote{If the emission point lies early enough, then some or all
rays will escape to infinity; on them $R$ need not have extrema. An example is
the ray marked ``out'' in the left panel of Fig. \ref{drawsecondproj}.} The
earliest emission point has $t = 5.075$, the later ones are $\Delta t = 0.0014$
apart, the last one at $t = 5.096$ is close below the $R = 2M$ hypersurface
which this observer would cross at $t = 5.0962668$. For each emission point
there are 512 rays emitted in initial directions inclined by $\pi / 256$ to each
other. For more details of this family of rays see Appendix \ref{M012}.


The left panel of Fig. \ref{duzeringi} shows the projections of loci of $R$
maxima on a surface of constant $t = t_0$. (In comoving coordinates, the
projection does not depend on $t_0$. This is not an isometric image because a
\{t = constant, $\vartheta = \pi/2$\} surface in an L--T model is not flat when
$E \neq 0$, see (\ref{2.1}).) There are no $R$ minima for these emission points,
and on rays which go off the initial point with $k_o^r \leq 0$ there are no
maxima either. The ring of $R$ maxima for the latest emission point coincides
with the center of the cross at the scale of the figure. The locus of all $R$
maxima is in this case a curved cone with the vertex at the intersection of the
observer world line with the $R = 2M$ hypersurface. As predicted, all maxima
occur at $R \geq 2M$ -- the right panel of Fig. \ref{duzeringi} shows this.

We considered rays running in the $\vartheta = \pi/2$ hypersurface, but in view
of comment (\ref{3.8}) this is not a great limitation. The whole bundle of rays
emitted from a fixed initial point consists of sub-bundles, each of which
contains rays running in a different $\vartheta' = \pi/2$ hypersurface where
$\vartheta'$ is related to $\vartheta$ by a rotation around a point. So, the
complete projection of the whole set of $R$ maxima on a 3-dimensional space of
constant $t$ can be imagined by rotating the left panel of Fig. \ref{duzeringi}
around the $\varphi = 0$ semiaxis.

Figure \ref{drawsecondproj} shows the projections of odd-numbered rings 1,
$\dots$, 15 of $R$ maxima on the $\varphi = 0$ surface (left panel) and on the
$\varphi = \pi / 2, 3 \pi / 2$ surface (right panel, horizontal scale smaller
than in Fig. \ref{duzeringi}). The rings are not plane curves. The intersections
of the lines in the left panel are artifacts of the projection; the only true
points of contact between $R = 2M$ and the maximum $R$ rings are on radial rays.
In the right panel, the continuous lines are intersections of the $R = 2M$
surface with the planes of constant $x = M \cos \varphi$.

\begin{figure}[h]
 ${}$ \\[5mm]
 \hspace{-5mm} \includegraphics[scale = 0.8]{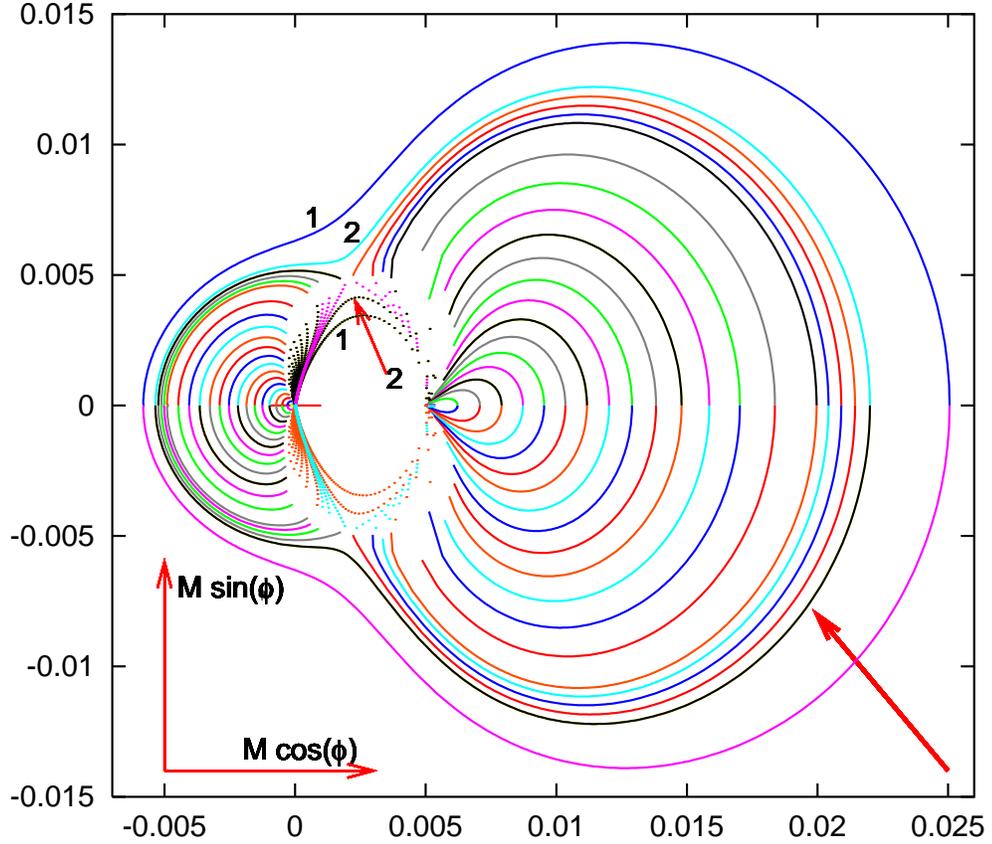}
\caption{The loci of $R$ maxima (continuous lines) and minima (dotted lines)
along rays going off several emission points on the $(M, \varphi) = (0.005, 0)$
emitter world line, projected on a $\{t = {\rm constant}, \vartheta = \pi/2\}$
surface along the dust flow lines. See the text for details, and Fig.
\ref{drawallextreflatlupa} for an enlarged view of the region around the tip of
the small arrow.}
 \label{drawallextreflat2}
\end{figure}

\begin{figure}[h]
 ${}$ \\[-1cm]
\begin{center}
 \hspace{-4cm} \includegraphics[scale = 0.55]{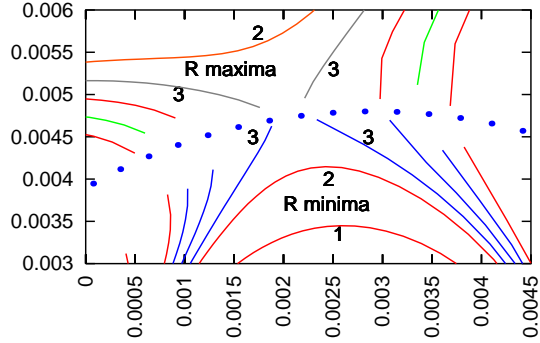}
\caption{A closeup view of the region in Fig. \ref{drawallextreflat2} where the
loops of $R$ extrema change geometry. The arc of large dots separates the loci
of maxima from the loci of minima.}
 \label{drawallextreflatlupa}
\end{center}
\end{figure}

The locus of $R$ extrema in Figs. \ref{duzeringi} and \ref{drawsecondproj} has a
simple shape because the emitter world line at $M = 0.012$ is far from the
center and the earliest emission point is sufficiently late. The geometry of
this locus is more complicated when the comoving emitter is closer to $M = 0$.
Consider the extrema of $R$ along bundles of future-directed rays emitted at
$(M, \varphi) = (0.005, 0)$, still in the $\vartheta = \pi/2$ hypersurface.
Here, as the emission instant progresses toward the future, the contours of $R$
extrema undergo an interesting evolution illustrated in Figs.
\ref{drawallextreflat2} and \ref{drawallextreflatlupa}. They show the
projections (along the cosmic dust flow lines) of the loci of $R$ extrema on a
$t = {\rm constant}$ surface, for rays going off 21 initial points. In the main
sequence of 18 emission points their $t$ coordinates change from $t = 5.05$ at
steps of 0.0025 to $t = 5.0925$. The last point is just below the $R = 2M$
surface. In addition, there are 3 emission points with $t = 5.053125 + j \times
0.000625$, where $j = 0, 1, 2$; the rays emitted at them allow for a more
detailed view of the evolution of the contours.

\begin{figure}[h]
\begin{center}
 ${}$ \\[-2.5cm]
 \includegraphics[scale = 0.55]{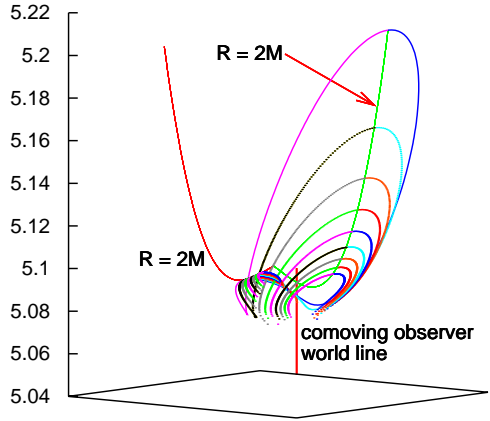}
 ${}$ \\[-2cm]
\caption{A 3d view of the loci of $R$ extrema on contours 1, 2, 6, $\dots$ ,
12.}
 \label{proj3d}
\end{center}
\end{figure}

At each initial point 512 rays were emitted, in regularly spaced initial
directions just as before. This time, along some rays $R$ has both minima and
maxima. In Fig. \ref{drawallextreflat2} the loci of maxima are the continuous
lines, the loci of minima are the dots. As the emission instant $t$ progresses,
the loops at left shrink toward the center of symmetry of the space at $M = 0$.
The loops at right shrink toward the emitter world line. The long arrow marks
the view direction in Fig. \ref{proj3d} (40$^{\circ}$ counterclockwise from the
$\varphi = 3 \pi/2$ semiaxis).

For the two earliest emission points, $R$ has a maximum along every ray, and a
minimum along rays going off at sufficiently large angles to the $\varphi = 0$
line. The contours of minima are initially inside the contours of maxima, but
approach each other with progressing emission instant (contours 1 and 2 in Figs.
\ref{drawallextreflat2} and \ref{drawallextreflatlupa}). Near to emission
instant 1 of the additional sequence, the contours come into contact (curves 3
in Fig. \ref{drawallextreflatlupa}). For later emission times the extrema again
form two disjoint loops, but each one is outside the other and contains both
maxima and minima. Up to emission instant 9 of the main sequence, the Fortran
program found at least two $R$ minima on the right-hand loops. For later
emission times, $R$ minima exist only on the left-hand loops. On rays that run
between the loops $R$ decreases monotonically to 0 at the BC.

As before, in view of comment (\ref{3.8}), also here the projection of all $R$
extrema on a $t =$ constant space can be imagined by rotating Fig.
\ref{drawallextreflat2} around the $\varphi = 0$ semiaxis.

The contours of $R$ extrema are again not plane curves, Fig. \ref{proj3d} shows
a 3d view of a few of them; they all lie in the $R \geq 2M$ region. The viewing
direction is at 85$^{\circ}$ to the $t$ axis and at 40$^{\circ}$
counterclockwise from the $\varphi = 3 \pi/2$ half-plane.

\begin{figure}[h]
 \begin{center}
 \includegraphics[scale = 0.7]{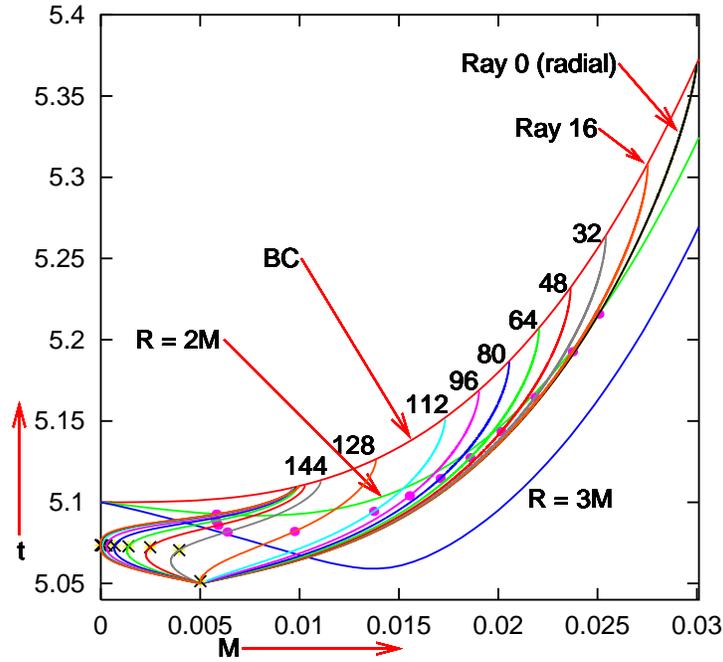}
\caption{The graphs of $t(M)$ along selected rays originating at $(M, t) =
(0.005, 5.05)$ in the L--T model defined by (\ref{5.1}) -- (\ref{5.3}). The dots
mark the $(M, t)$ coordinates of the maxima of $R$. On Rays 128, $\dots$ 256,
$R$ has also minima marked by $\times$s. Note that $2M \leq R < 3M$ for all
maxima, and $R > 3M$ for all minima. }
 \label{drawtM}
 \end{center}
\end{figure}

In order to further visualise the conclusions of Sec. \ref{Rextre}, Fig.
\ref{drawtM} shows the $t(M)$ profiles along selected rays of the bundle that
created contour 1 in Fig. \ref{drawallextreflat2}. Rays beyond \# 144 have
numbers $160 + j \times 16$ with $j = 0, \dots, 6$; their labels are omitted.
The dots mark the $(M, t)$ coordinates of the loci of $R$ maxima along the rays;
at each one $2M \leq R < 3M$. The rightmost and leftmost rays are radial, and
along them $R$ is maximum where $R = 2M$. Minima of $R$ exist only along Rays
128 and following, their loci are marked by $\times$s. At each minimum $R > 3M$.

\begin{figure}[h]
\begin{center}
 ${ }$ \\ [1cm]
 \hspace{-3cm} \includegraphics[scale = 0.7]{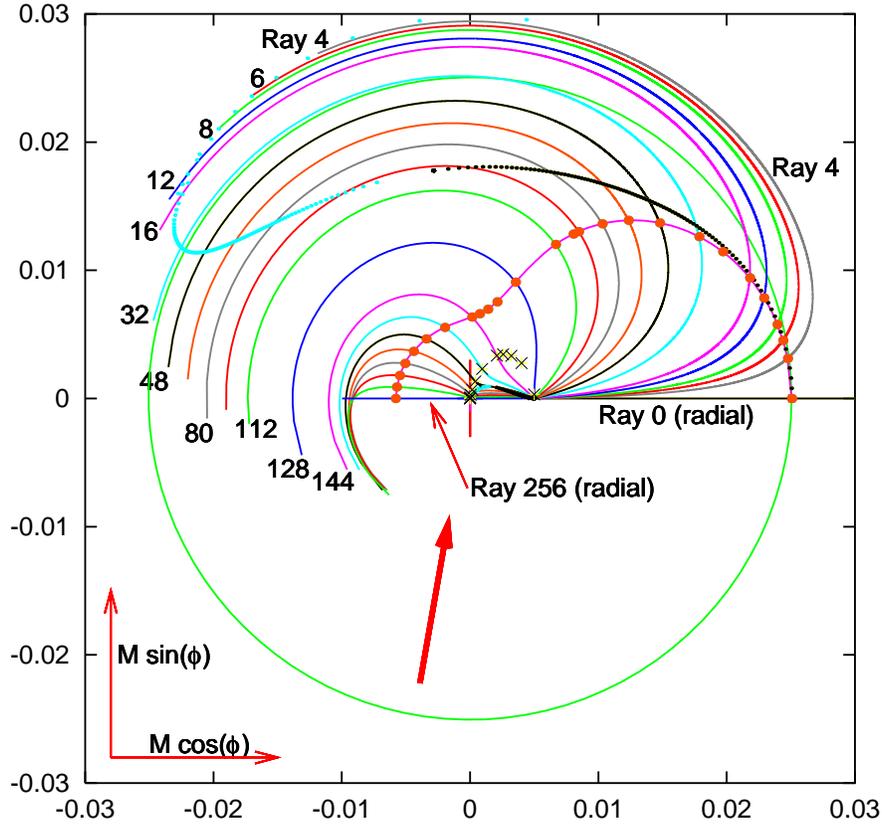}
\caption{Projections of selected rays of the earliest bundle in Fig.
\ref{drawallextreflat2} on a surface of constant $t$ along the flow lines of the
cosmic dust. See the text for explanations. }
 \label{drawrecxy}
\end{center}
\end{figure}

\begin{figure}[h]
\begin{center}
 ${}$ \\[1mm]
 \hspace{3cm} \includegraphics[scale = 0.8]{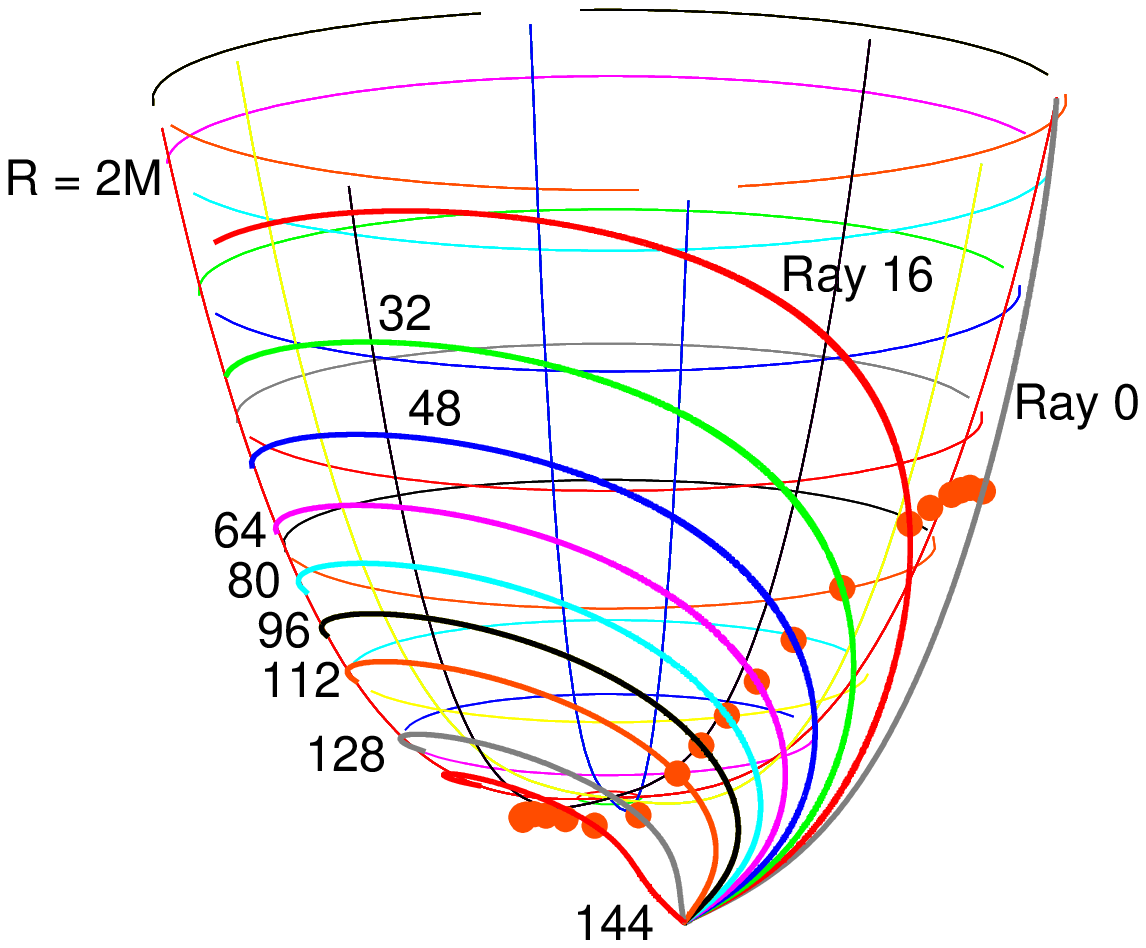}
 ${}$ \\[-4.2cm]
 \hspace{-17cm} \includegraphics[scale = 0.4]{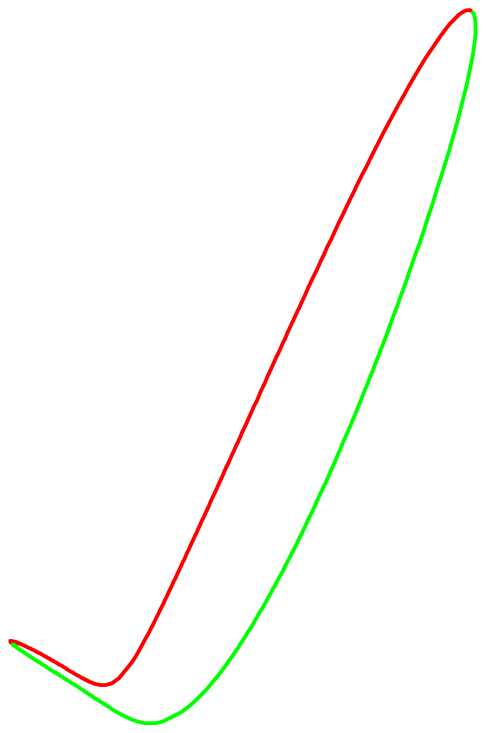}
 ${}$ \\[-1cm]
\caption{{\bf Right graph:} A 3d view of selected rays from Figs. \ref{drawtM}
and \ref{drawrecxy} in the $(M \cos \varphi, M \sin \varphi, t)$ coordinate
space. The paraboloid-like surface is the $R = 2M$ locus. {\bf Left graph:} A 3d
view of the full ring of maxima of $R$ along rays emitted at $(M, \varphi, t) =
(0.005, 0.0, 5.05)$. See the text for details. }
 \label{drawtxy}
\end{center}
\end{figure}

Figure \ref{drawrecxy} shows the projections of selected rays of the
earliest-emitted bundle in Fig. \ref{drawallextreflat2} on a $t = $ constant
surface along the flow lines of the cosmic dust. The large dots and the
$\times$-s mark the $(x, y) \df (M \cos \varphi, M \sin \varphi)$ coordinates of
the points where $R$ is maximum and, respectively, minimum (some extrema are
shown without the rays on which they occur). All the extrema lie along contour 1
of Fig. \ref{drawallextreflat2}. The large circle is at $R = 2M$ on the outward
radial Ray 0, where $M = 0.025058$. Each curve ends just before the ray would
cross the BC. The vertical stroke marks $M = 0$. The thick arrow marks the view
direction in Fig. \ref{drawtxy}. The meaning of the curve of small dots will be
explained in Sec. \ref{exemzerotheta}.


Figure \ref{drawrecxy} visualises one more fact: the transition from nonradial
to radial rays is discontinuous. Nonradial rays meet the singularity
tangentially to $r =$ constant rings while the radial ones meet the singularity
orthogonally to those rings.

The right-hand graph in Fig. \ref{drawtxy} shows a 3-d image of selected rays
from Figs. \ref{drawtM} and \ref{drawrecxy}, and also of the $R = 2M$
hypersurface. It is a map of the $\vartheta = \pi/2$ subspace of the L--T model
(\ref{5.1}) -- (\ref{5.3}) into a Euclidean space with coordinates $(x, y, t) =
(M \cos \varphi, M \sin \varphi, t)$. The loop at left is the full ring of
maxima of $R$, i.e., it includes the $x < 0$ half of the bundle (omitted in Fig.
\ref{drawrecxy}). The loci of $R$ maxima marked by large dots in the right graph
lie along the left half of the loop, which is further from the viewer. The
viewing direction in both graphs, marked by the arrow in Fig. \ref{drawrecxy},
is at 85$^{\circ}$ to the vertical axis and at 10$^{\circ}$ clockwise from the
$\varphi = 3 \pi/2$ half-plane.

\section{The set ${k^{\mu}};_{\mu} = 0$ in a general L--T model}\label{extexp}

\setcounter{equation}{0}

For rays with tangent vectors $k^{\mu}$ in the metric (\ref{2.1}) we have, using
(\ref{3.6})
\begin{eqnarray}\label{6.1}
&& \theta \df {k^{\mu}};_{\mu} \equiv \left(\sqrt{-g} k^{\mu}\right),_{\mu} /
\sqrt{-g} = \\
&& \frac {\sqrt{1 + 2E}} {R^2 R,_r \sin \vartheta} \left[\left(\frac {R^2 R,_r
\sin \vartheta k^t} {\sqrt{1 + 2E}}\right),_t + \left(\frac {R^2 R,_r \sin
\vartheta k^r} {\sqrt{1 + 2E}}\right),_r + \left(\frac {R^2 R,_r \sin \vartheta
k^{\vartheta}} {\sqrt{1 + 2E}}\right),_{\vartheta}\right]. \nonumber
\end{eqnarray}
When $C^2 \neq {J_0}^2 > 0$, (\ref{6.1}) becomes, using (\ref{3.7})
($\varepsilon_1 = \pm 1$ is the sign of $k^{\vartheta}$),
\begin{eqnarray}\label{6.2}
&& {k^{\mu}};_{\mu} = {k^t},_t + {k^r},_r + \left(2\ \frac {R,_t} R + \frac
{R,_{tr}} {R,_r}\right) k^t + \left(2\ \frac {R,_r} R + \frac {R,_{rr}} {R,_r} -
\frac {E,_r} {1 + 2E}\right) k^r \nonumber \\
&& + \frac {\varepsilon_1 C^2 \cos \vartheta} {R^2 \sqrt{C^2 \sin^2 \vartheta -
{J_0}^2}}.
\end{eqnarray}
When $C^2 = {J_0}^2 > 0$, (\ref{3.7}) implies $\vartheta = \pi/2$ and
$k^{\vartheta} = 0$. Then the $k^{\vartheta}$ term in (\ref{6.1}) disappears,
and the last term in (\ref{6.2}) does not arise. This case can be formally
included in (\ref{6.2}) using the convention that the limit of the last term at
${J_0}^2 \to C^2$ is zero.

Using (\ref{3.11}) and $\dril {k^r} {\lambda} = (\dril {k^r} t) k^t + (\dril
{k^r} r) k^r$, Eq. (\ref{3.2}) becomes
\begin{eqnarray}\label{6.3}
&& k^t {k^r},_t + k^r {k^r},_r + 2\ \frac {R,_{tr}} {R,_r}\ k^t k^r +
\left(\frac {R,_{rr}} {R,_r} - \frac {E,_r} {1 + 2E}\right) \left(k^r\right)^2 -
\frac {C^2 (1 + 2E)} {R^3 R,_r} = 0.\ \ \ \ \ \
\end{eqnarray}
Now, we differentiate by $t$ the null condition (\ref{3.12}):
\begin{equation}\label{6.4}
k^t {k^t},_t - \frac {R,_r R,_{tr}} {1 + 2E}\ \left(k^r\right)^2 - \frac
{{R,_r}^2} {1 + 2E}\ k^r {k^r},_t + \frac {C^2 R,_t} {R^3} = 0.
\end{equation}
We multiply (\ref{6.3}) by ${R,_r}^2 k^r/(1 + 2E)$, (\ref{6.4}) by $k^t$ and
add. The result is
\begin{eqnarray}\label{6.5}
&& \frac {\left(R,_r k^r\right)^2} {1 + 2E} \left[{k^r},_r + \frac {R,_{tr}}
{R,_r}\ k^t + \left(\frac {R,_{rr}} {R,_r} - \frac {E,_r} {1 + 2E}\right)
k^r\right] \nonumber \\
&& + \frac {C^2} {R^3} \left(R,_t k^t - R,_r k^r\right) + \left(k^t\right)^2
{k^t},_t = 0.\ \ \ \ \
\end{eqnarray}
We now calculate ${k^t},_t$ from (\ref{6.5}) and substitute it in (\ref{6.2})
obtaining
\begin{eqnarray}\label{6.6}
&&{k^{\mu}};_{\mu} = 2\ \frac {R,_t} R\ k^t + 2\ \frac {R,_r} R\ k^r + \frac
{\varepsilon_1 C^2 \cos \vartheta} {R^2 \sqrt{C^2 \sin^2
\vartheta - {J_0}^2}} \nonumber \\
&& + \frac {C^2} {\left(k^t\right)^2 R^2} \left[- \frac {R,_t k^t - R,_r k^r} R
+ {k^r},_r + \frac {R,_{tr}} {R,_r} k^t + \left(\frac {R,_{rr}} {R,_r} - \frac
{E,_r} {1 + 2E}\right) k^r\right].
\end{eqnarray}

For the solutions of ${k^{\mu}};_{\mu} = 0$ in the collapse phase of the model
there are four cases:

(1) For outward radial rays $C = 0$, so ${k^{\mu}};_{\mu} = 2 \left(R,_t k^t +
R,_r k^r\right)/R$. Then the loci of $\theta = 0$ and of maximum $R$ coincide
and are at $R = 2M$ \cite{PlKr2006}.

(2) For inward radial rays $\theta \neq 0$ all along. This follows from
(\ref{6.6}): on radial rays $C = 0$, so $\theta = 0$ would coincide with $R,_t
k^t + R,_r k^r = 0$. But on an inward ($k^r < 0$) future-directed ($k^t > 0$)
ray in the collapse phase ($R,_t < 0$) we have $R,_t k^t + R,_r k^r < 0$ all
along because $R,_r > 0$ (no shell crossings).\footnote{However, it may happen
that an inward radial ray flies through the center $M = 0$ and becomes outward
on the other side. On the outward segment, point (1) applies. See examples in
Secs. \ref{Rextreexam} and \ref{exemzerotheta}.}

(3) On nonradial rays ($C \neq 0$), $R,_t k^t + R,_r k^r = 0$ does not fulfil
${k^{\mu}};_{\mu} = 0$ identically, so the locus of $\theta = 0$ in general does
not coincide with the locus of extrema of $R$ (but see footnote \ref{noconfu} in
Sec. \ref{intro}; see also Figs. \ref{drawtimes} and \ref{drawtimesmale} for
exceptions). Then

(3a) For rays with $J^2 = C^2$, which stay in $\vartheta = \pi/2$, solutions of
${k^{\mu}};_{\mu} = 0$ (when they exist) determine a curve in a $(t, r)$
surface, and a surface in a $(t, r, \varphi)$ space.

(3b) When $C^2 > {J_0}^2$, solutions of ${k^{\mu}};_{\mu} = 0$ determine a
2-surface in the $(t, r, \vartheta)$ space, and the locus of $\theta = 0$ is a
3-dimensional subspace of the L--T spacetime.

In case (3) the derivative ${k^r},_r$ in (\ref{6.6}) goes across the bundle. To
calculate it numerically two rays are needed: a $G_1$ with a given $C_1/R_o$,
and a nearby $G_2$ with $C_2/R_o \df$ $C_1/R_o + D$. It must be calculated at
constant $t, \vartheta$ and $\varphi$, so for each point $p_1$ on $G_1$ we find
the point $p_2$ on $G_2$ with the same $(t, \vartheta, \varphi)$, where $r =
r_2$ and $k^r = k_2^r$. Then
\begin{equation}\label{6.7}
{k^r},_r \approx \frac {k^r_2 - k^r_1} {r_2 - r_1}.
\end{equation}
All other quantities in (\ref{6.6}) are intrinsic to a single geodesic. In fact,
nothing depends on $\varphi$ in (\ref{6.6}), and in case (3a) nothing depends on
$\vartheta$ either; then it suffices to find a point on $G_2$ with the same $t$.
For more comments on (\ref{6.7}) see Appendix \ref{numerics}.

At such points where $r_2 = r_1$ but $k^r_2 \neq k^r_1$, $\left|{k^r},_r\right|
\to \infty$ and may jump between $\pm \infty$. This may be a real effect or a
numerical artifact. Note that a real jump of ${k^r},_r$ may only be from $-
\infty$ to $+ \infty$, and the same is true for $\theta$ -- see Appendix
\ref{jump} for a proof. This has a geometrical interpretation: the jump from
$\theta = - \infty$ to $\theta = + \infty$ means that the ray bundle was
refocussed to a point and then disperses; the opposite is hard to imagine.

In case (3b), for each initial point and each given $C$, one has to consider the
bundle of rays with the same $C$ and all $J_0$ allowed by (\ref{3.9}). A
graphical representation of such an object would be a problem in itself, so, for
this introductory study, we shall consider case (3a) only. But, in view of
remark (\ref{3.8}), in this way we disallow only the auxiliary nearby rays with
${J_0}^2 < C^2$ because the $(\vartheta, \varphi)$ coordinates can be adapted to
each sub-family of the main rays that proceed in the same equatorial
hypersurface.

\section{The set ${k^{\mu}};_{\mu} = 0$ in the exemplary L--T model of Sec.
\ref{Rextreexam}}\label{exemzerotheta}

\setcounter{equation}{0}

\begin{figure}
\begin{center}
 \includegraphics[scale=0.7]{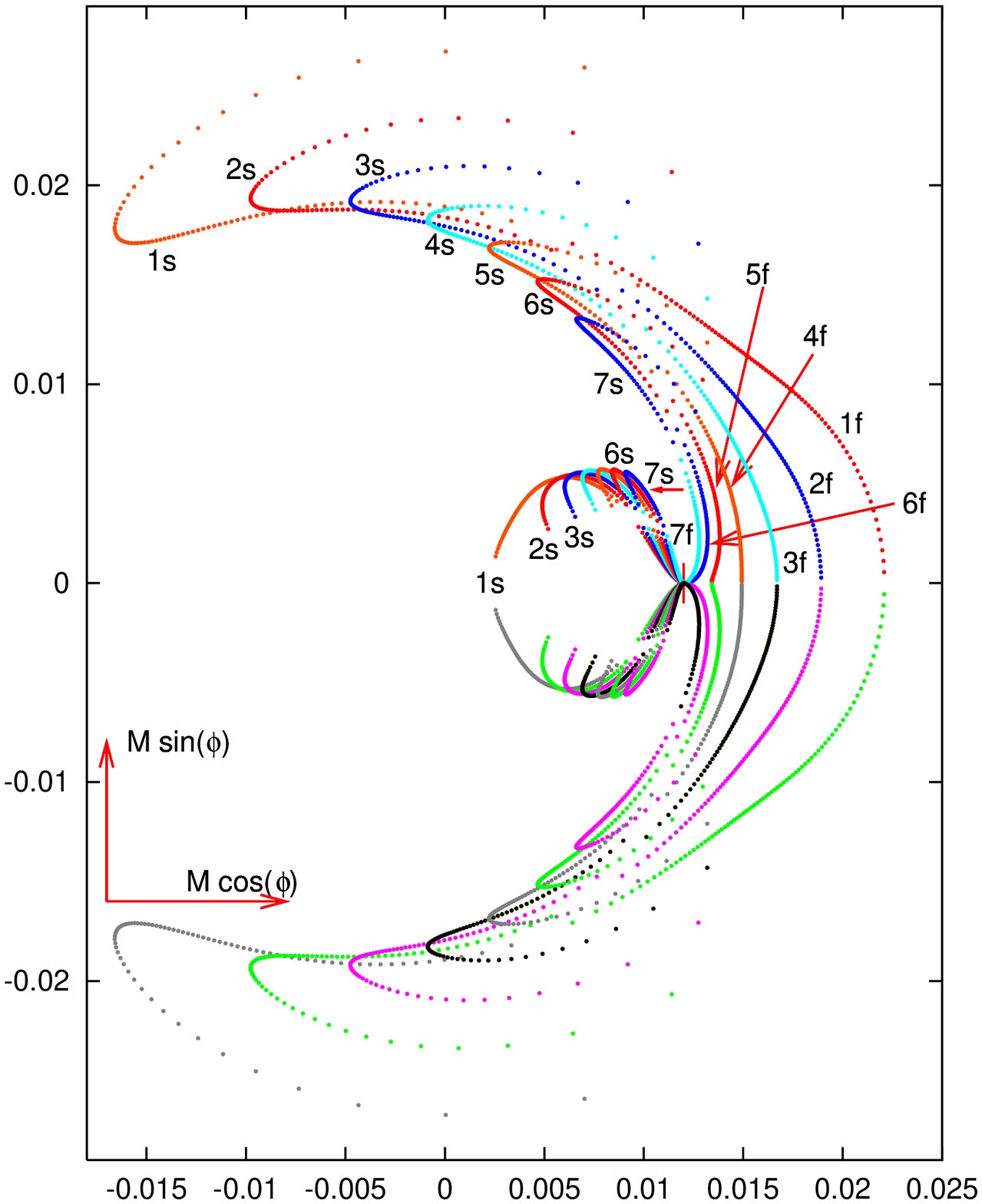}
 ${}$ \\[-11cm]
 \hspace{-4.8cm} \includegraphics[scale=0.6]{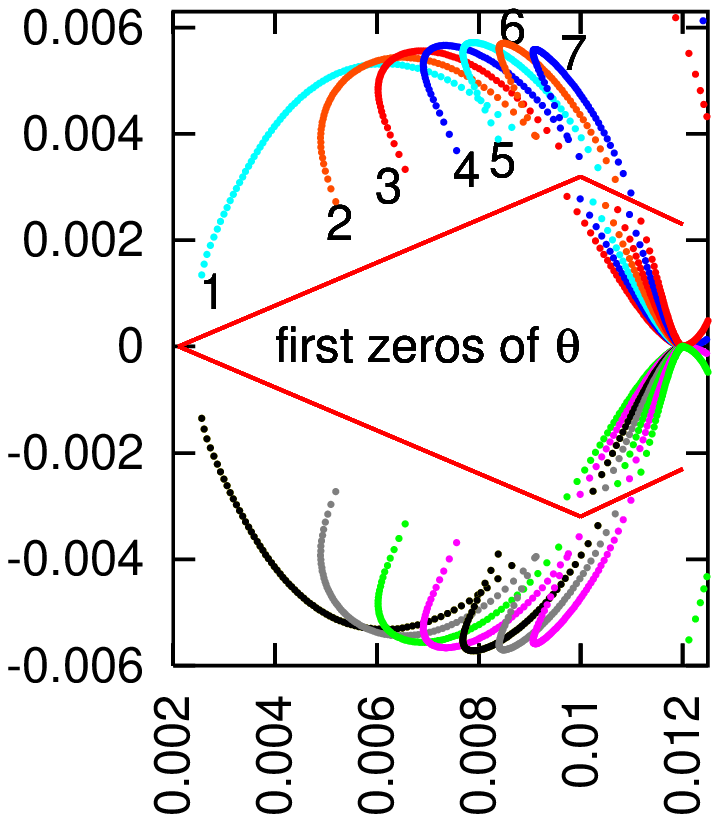}
 ${}$ \\[6.5cm]
\caption{{\bf Main panel:} Loci of $\theta = 0$ for rays emitted at 7 selected
instants on the world line $(M, \varphi) = (0.012, 0)$ (marked by the vertical
stroke) in the model (\ref{5.1}) -- (\ref{5.3}), projected on a surface of
constant $t$. {\bf Inset:} A closeup view on the central blob. Labels refer to
the emission instants. See the text for explanations. }
 \label{drawzeradrugieM}
\end{center}
\end{figure}

\begin{figure}[h]
 ${}$ \\[-1cm]
\begin{center}
\includegraphics[scale=0.65]{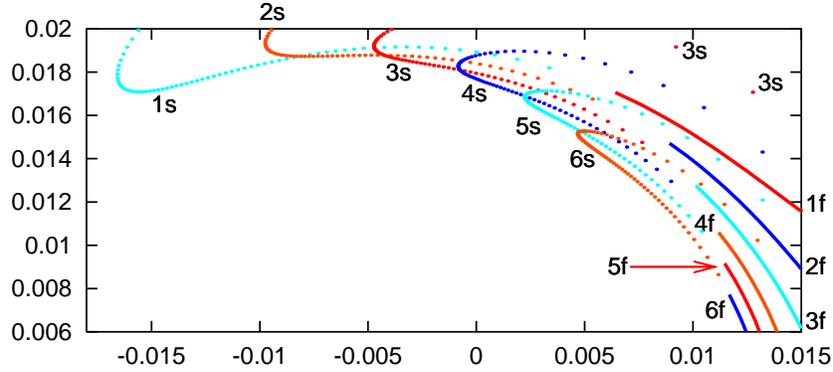}
\caption{A closeup view on the upper part of Fig. \ref{drawzeradrugieM}. }
 \label{drawzeradrugieMlupa}
\end{center}
\end{figure}

We first consider bundles of rays going off the same observer world line $(M,
\varphi) = (0.012, 0)$ as in Figs. \ref{duzeringi} and \ref{drawsecondproj} --
see Figs. \ref{drawzeradrugieM} and \ref{drawzeradrugieMlupa}. The origins of
the bundles are here at points 1, 4, 7, 10, 13 and 16 of the former set (here
labelled 1, $\dots$, 6) and the additional point 7 at $(M, t) = (0.012,
5.1002)$. At point 7 $R > 2M$ (for this observer $R = 2M$ is at $t = 5.09627$).
In each bundle there are 512 main rays distributed in the same way as before.
Let $C/R_o = d_1$ for the main ray and $d_2$ for the auxiliary ray used to
calculate ${k^r},_r$. The auxiliary rays have $d_2 = d_1 + 1/1024$ for main Rays
1 -- 127 and $d_2 = d_1 - 1/1024$ for main Rays 128 -- 255 (so each auxiliary
ray goes off at a larger angle to the $\varphi = 0$ semiaxis than the main ray.)
The loci of $\theta = 0$ for Rays 257 -- 511 are found by inverting the $y = M
\sin \varphi$ coordinates of those on Rays 1 -- 255.


\begin{figure}[h]
 ${}$ \\[-1cm]
\begin{center}
\includegraphics[scale=0.6]{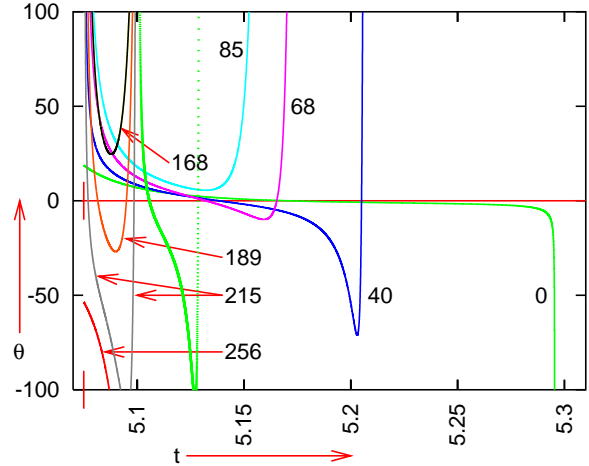}
\caption{The graphs of $\theta(t)$ along selected rays emitted at $(M, t) =
(0.012, 5.075)$ (the continuous curves) and $(0.012, 5.1002)$ (the dotted
curve). }
 \label{drawduzeth0}
\end{center}
\end{figure}

For each emission point, $\theta$ has no zeros along several rays. For example,
for point 1 at $t = 5.075$, there are no $\theta$ zeros on Rays 79 -- 178 (and
on their mirror-images 334 -- 433) and on the inward radial ray. On each outward
radial ray, $\theta$ has a single zero. On most remaining rays $\theta$ has two
zeros. On each boundary ray between those with two zeros and those with no zeros
$\theta$ has a single zero. The meaning of the symbols in Figs.
\ref{drawzeradrugieM} and \ref{drawzeradrugieMlupa} is: 1f = initial point 1,
first zero of $\theta$, 1s = initial point 1, second zero of $\theta$, 2f =
initial point 2, first zero, etc. The inset in Fig. \ref{drawzeradrugieM} shows
the central blob enlarged; the numbers in it identify the emission points. Rays
with no $\theta$ zeros run between the blob and the long dotted arcs. Figure
\ref{drawzeradrugieMlupa} shows where the arcs of first zeros (continuous lines)
go over into the arcs of second zeros (dotted lines), each single zero lies at
their contact.

Figure \ref{drawduzeth0} shows the graphs of $\theta(t)$ along selected rays
emitted at point 1 and along Ray 40 emitted at point 7 (the dotted line). There
is a discontinuity between Ray 0 (on which $\theta$ has one zero) and the first
nonradial ray, on which $\theta$ has two zeros. Then the changes proceed
continuously up to Ray 255. Rays with a single $\theta = 0$ are between 78 and
79, and again between 178 and 179. There is one more discontinuity between the
last nonradial ray and the inward radial Ray 256, on which $\theta < 0$ all
along. The vertical strokes mark the $t = 5.075$ coordinate of point 1. The
dotted line shows that the $\theta(t)$ profile is still similar when the
emission point is in the $R < 2M$ region. On all rays except the two radial
ones, $\theta$ becomes very large {\em positive} on approaching the BC, which
means that the ray bundles diverge there. On those nonradial rays where $\theta$
has no zeros, the bundle is diverging all the time ($\theta > 0$, but not
monotonic, see graphs 85 and 168).

\begin{figure}[h]
 ${}$ \\ [-1cm]
\begin{center}
\includegraphics[scale=0.6]{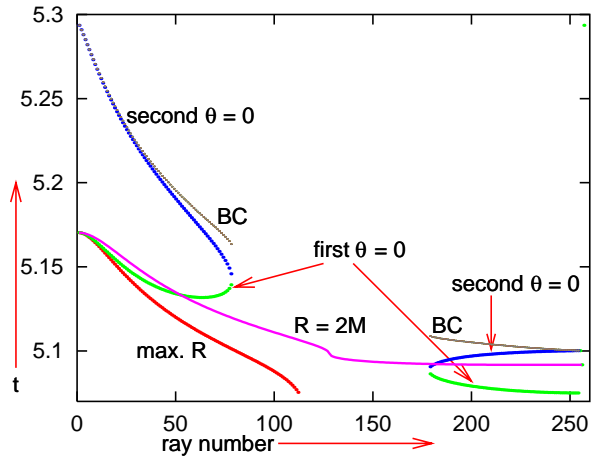}
\caption{The $t$ coordinates of various events on rays emitted at $(M, t) =
(0.012, 5.075)$. }
 \label{drawtimes}
\end{center}
\end{figure}

\begin{figure}[h]
\begin{center}
 ${}$ \\ [-1cm]
\includegraphics[scale=0.6]{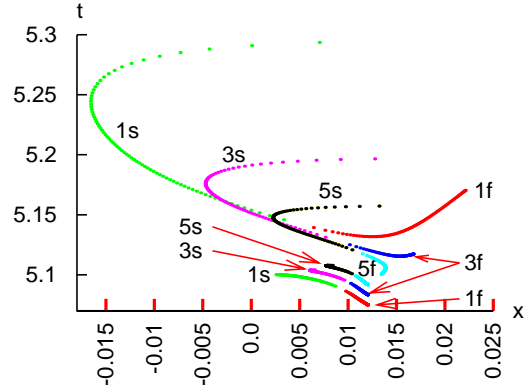}
 ${}$ \\ [-1cm]
\caption{The $\theta = 0$ curves corresponding to emission points 1, 3 and 5
from Fig. \ref{drawzeradrugieM} projected on the $y = M \sin \varphi = 0$
surface. }
 ${}$ \\ [-1cm]
 \label{drawduzeth03dfront}
\end{center}
\end{figure}

Figure \ref{drawtimes} shows the $t$ coordinates of the loci of maximum $R$, of
both $\theta = 0$ and of $R = 2M$ on all 256 rays emitted at point 1 with
$k^{\varphi} \geq 0$. The ray-number $j$ is related to the angle $\alpha_j$
between the initial direction of the ray and the $\varphi = 0$ semiaxis by
$\alpha_j = j \pi / 256$. The $\theta = 0$ curves intersect the $R = 2M$ curve
at two points, so there exist nonradial rays on which one locus of $\theta = 0$
is at $R = 2M$. The dotted curve marked "BC" is the graph of $t$ at BC at that
$M$ where the second $\theta = 0$ occurred on the ray ({\em not to be confused}
with $t$ at which the ray hits the BC!). It demonstrates that the locus of the
second $\theta = 0$ approaches the BC when $\alpha_j \to 0, \pi$, so the loose
ends of the dotted curves in Fig. \ref{drawzeradrugieM} are near the BC. The
time-ordering of $\theta = 0$ and $R = 2M$ in Fig. \ref{drawtimes} changes from
ray to ray. This has physical consequences, to which we will come back in Sec.
\ref{sumcon}.

\begin{figure}[h]
\begin{center}
 ${}$ \\ [1cm]
 \includegraphics[scale=0.7]{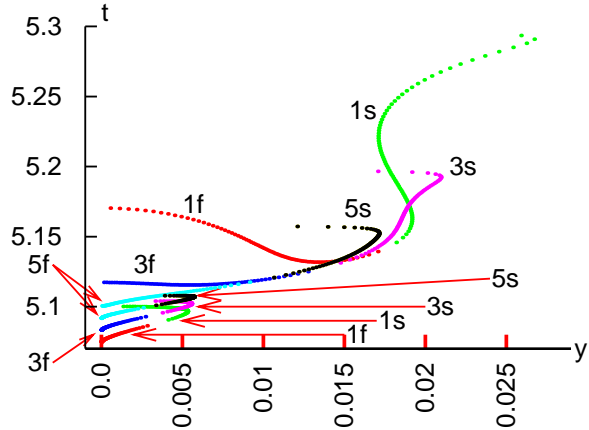}
 ${}$ \\ [-1.5cm]
\caption{The $\theta = 0$ curves corresponding to emission points 1, 3 and 5
from Fig. \ref{drawzeradrugieM} projected on the $x = M \cos \varphi = 0, y \geq
0$ half-plane.  }
 \label{drawduzeth03dside}
\end{center}
\end{figure}

\begin{figure}[h]
 ${}$ \\ [-2cm]
\begin{center}
 \hspace{5mm} \includegraphics[scale=0.7]{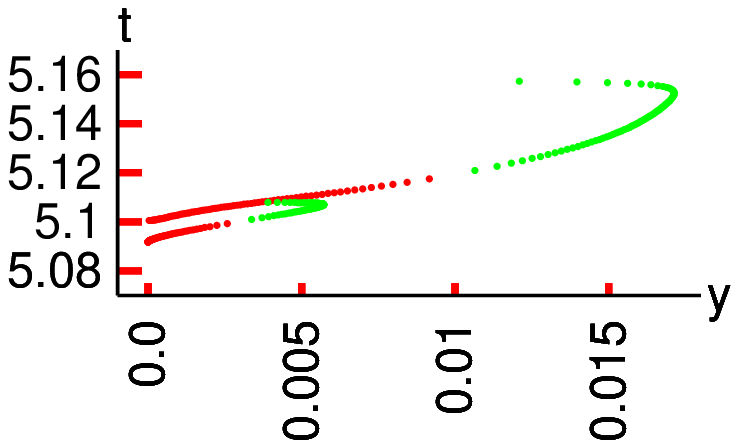}
 ${}$ \\ [-4.4cm]
\hspace{-12.5cm} \includegraphics[scale=0.7]{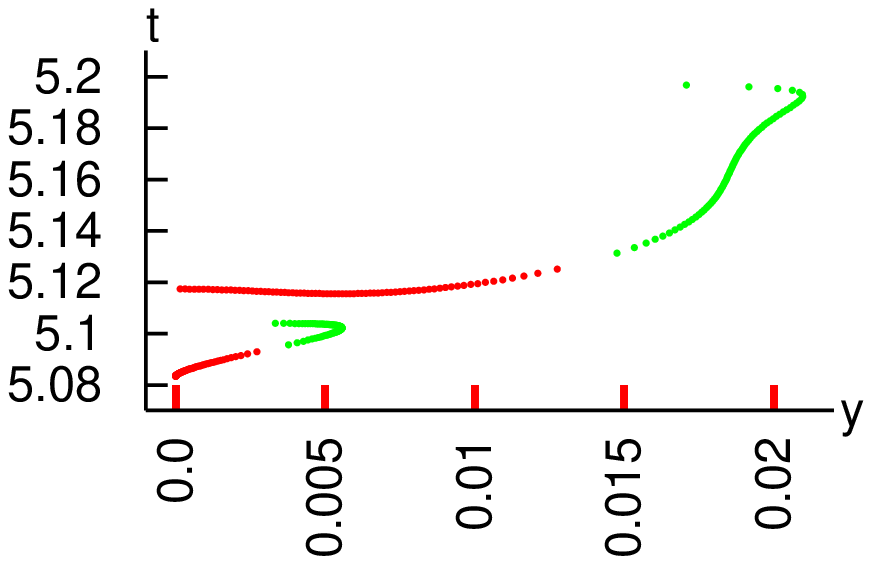}
 ${}$ \\ [-1cm]
\caption{The $\theta = 0$ curves for emission points 3 (at left) and 5 (at
right) from Fig. \ref{drawduzeth03dside}. The first $\theta$ zeros are to the
left of the gap in each curve, the second zeros are to the right. }
 \label{drawduzeth03dsidelupa}
\end{center}
\end{figure}

Figure \ref{drawduzeth03dfront} shows the projections of the $\theta = 0$ curves
corresponding to emission points 1, 3 and 5 on the $y = M \sin \varphi = 0$
surface. The upper ends of both branches of the 1s, 3s and 5s arcs are close to
the BC. Figure \ref{drawduzeth03dside} shows the $y \geq 0$ halves of the same
$\theta = 0$ curves as in Fig. \ref{drawduzeth03dfront}, this time projected on
$x = M \cos \varphi = 0$ surface. Since the three projections are somewhat
entangled, Fig. \ref{drawduzeth03dsidelupa} shows the curves for emission points
3 and 5 separately, at the same scale as in the previous figures.

Now we consider ray bundles emitted on the world line $(M, \varphi) = (0.005,
0)$. The instants of emission are at $t = 5.05 + k \times 0.0085$, where $k = 0,
1, \dots, 5$. (The latest and earliest emission points are the same as for the
ray bundles used in Fig. \ref{drawallextreflat2}.) At each of these instants,
512 rays are emitted in regularly spaced initial directions, just as before.


\begin{figure}[h]
\begin{center}
 ${}$ \\ [-6cm]
 \hspace {-8mm} \includegraphics[scale=0.7]{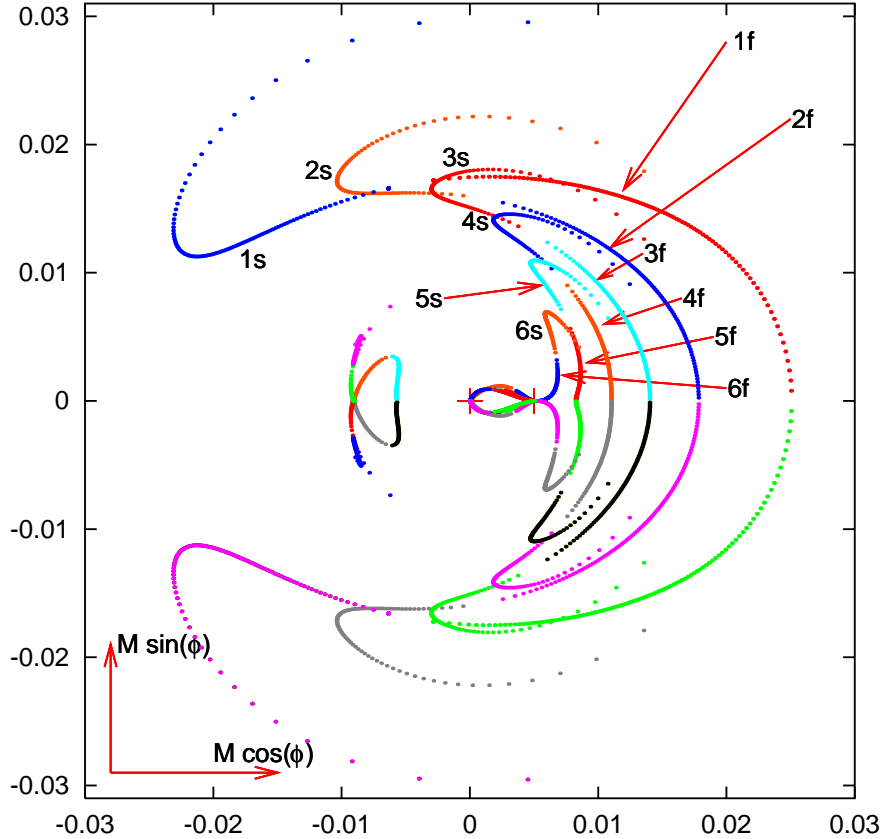}
\caption{The analogue of Fig. \ref{drawzeradrugieM} with the origins of the ray
bundles at $(M, \varphi) = (0.005, 0)$. The knot to the left of $M = 0$ is the
locus of 3rd, 4th, 5th and 6th $\theta$ zeros on the earliest bundle of rays.
See Fig. \ref{drawzeramaleMknot} and explanations in the text.}
 \label{drawzeramaleM}
\end{center}
\end{figure}

\begin{figure}[h]
\begin{center}
 ${}$ \\ [-10cm]
\includegraphics[scale=0.6]{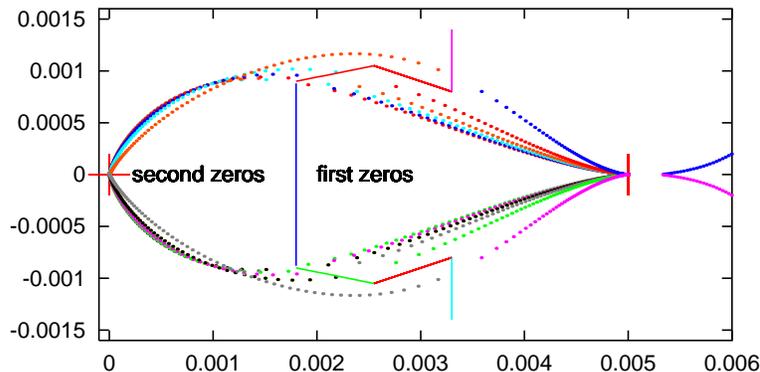}
\caption{A closeup view on the central blob in Fig. \ref{drawzeramaleM}. The
loop with the largest vertical diameter corresponds to emission point 6; the
vertical diameters decrease when proceeding from point 6 to point 1. }
 \label{drawzeramaleMlupa}
\end{center}
\end{figure}

\begin{figure}[h]
\begin{center}
 ${}$ \\ [-6cm]
\includegraphics[scale=0.7]{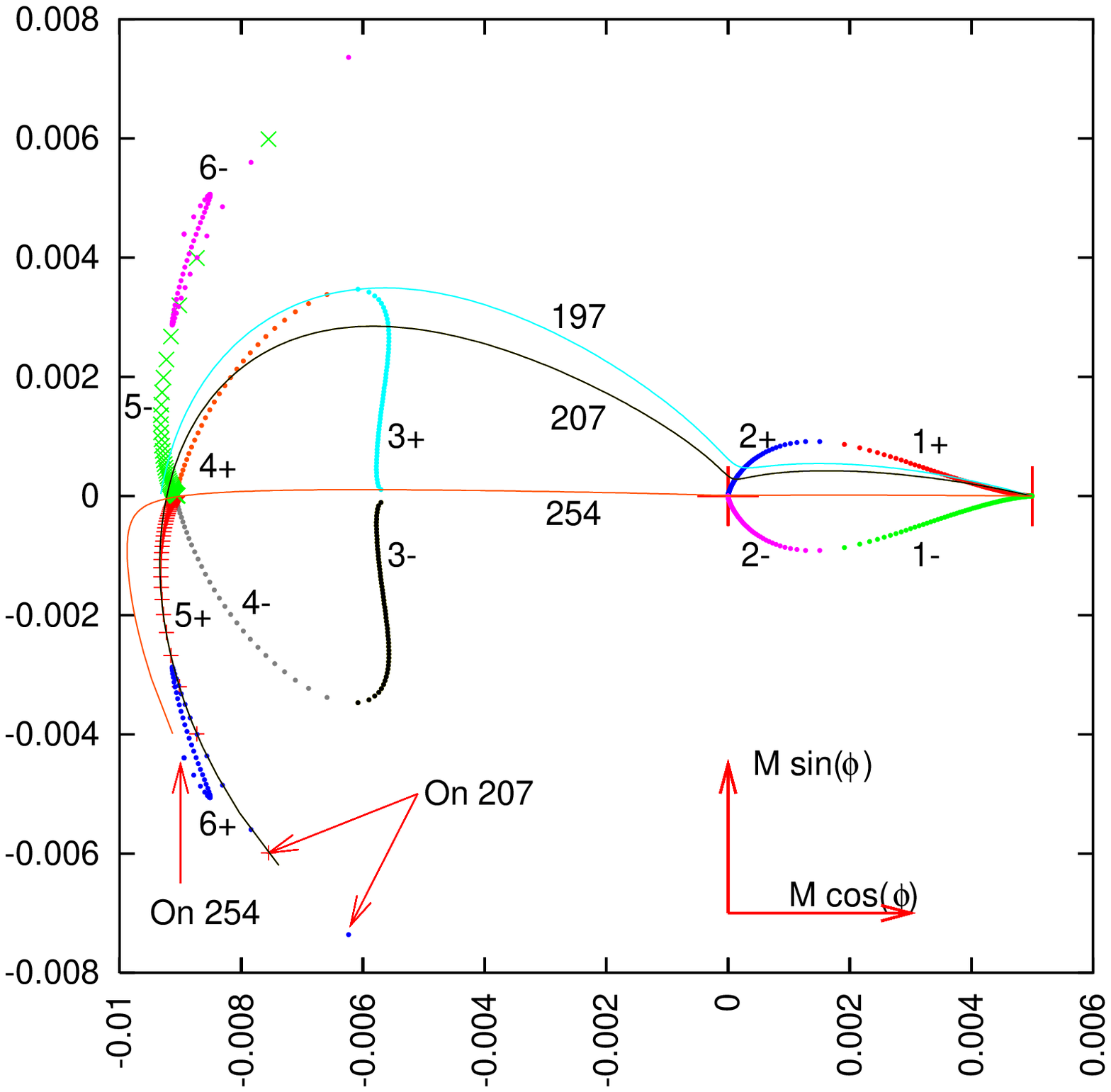}
\caption{The knot in Fig. \ref{drawzeramaleM} enlarged. See the text for
explanations.}
 \label{drawzeramaleMknot}
\end{center}
\end{figure}

\begin{figure}[h]
\begin{center}
\includegraphics[scale=0.6]{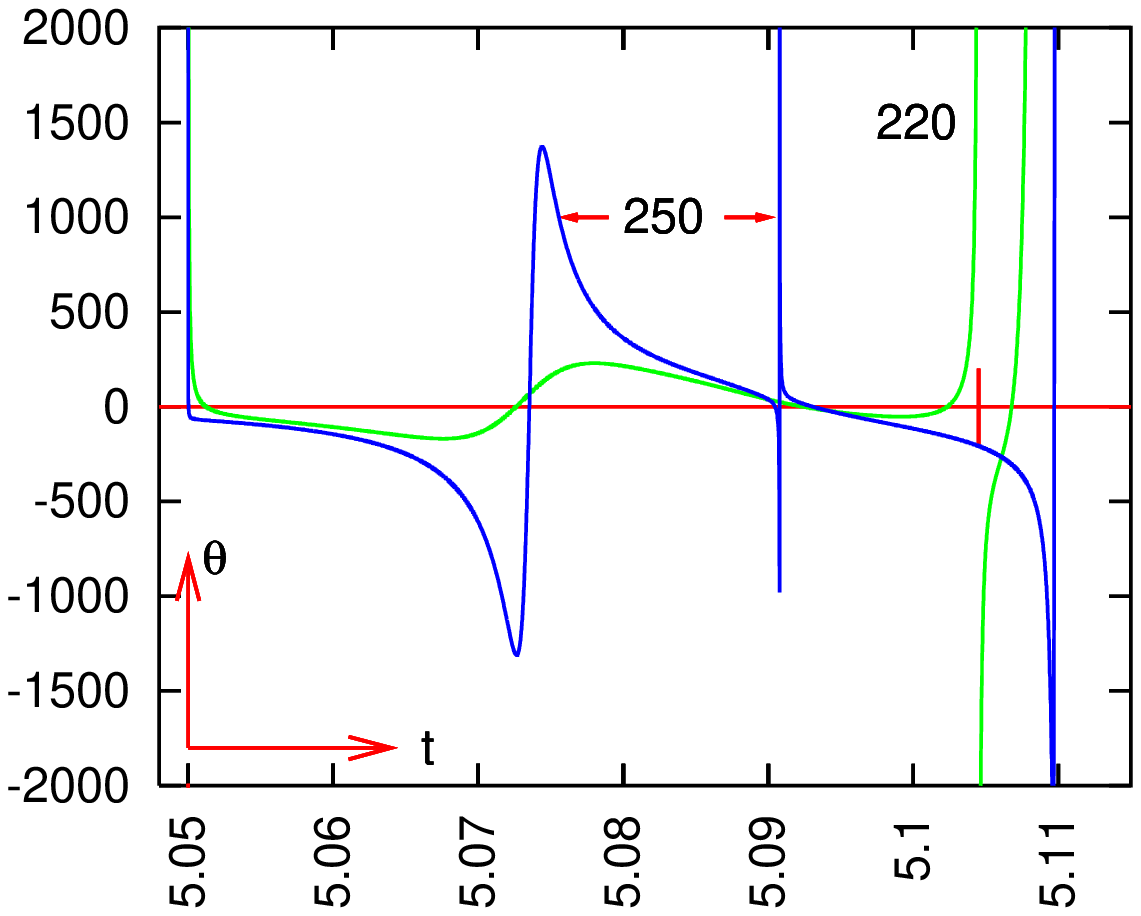}
\caption{Typical profiles of $\theta(t)$ on rays with 6 zeros. See remarks in
the text. }
 \label{drawmaleth06}
\end{center}
\end{figure}

This time, along some rays $\theta$ has 3 to 6 zeros. For example, in the bundle
emitted at point 1, $\theta$ has two zeros on Rays 1 -- 104 and 177 -- 196, no
zeros on Rays 105 -- 176, four zeros on Rays 197 -- 206 and six on Rays 207 --
254. Three zeros exist on a ray between 196 and 197 and 5 on one between 206 and
207. Ray 255 is different: its $\theta(t)$ profile is similar to curve 250 in
Fig. \ref{drawmaleth06}, except that it begins with $\theta < 0$, so it has 5
zeros. Ray 256 passes through $M = 0$ and becomes outward on the other side,
where $\theta$ has a single zero at $R = 2M$. The reason of more $\theta$ zeros
is that the emitter world line is now close to $M = 0$, so some rays fly near
the center and later recede from it.

Figure \ref{drawzeramaleM} shows the projections on a surface of constant $t$ of
the loci of the first two $\theta$ zeros on each ray for all emission points,
and the loci of zeros \# 3, $\dots$, 6 for emission point 1 (this is the knot
left of $M = 0$). The upper half of the largest contour is shown in small dots
in Fig. \ref{drawrecxy}. The projection of the emitter world line is marked with
the vertical stroke, the cross marks the center $M = 0$. The meaning of the
labels is the same as in Fig. \ref{drawzeradrugieM}. Figure
\ref{drawzeramaleMlupa} shows a closeup view on the central part of Fig.
\ref{drawzeramaleM}.

Figure \ref{drawzeramaleMknot} shows the knot in Fig. \ref{drawzeramaleM}
enlarged. This is the collection of loci of $\theta$ zeros \# 3, $\dots$, 6 for
emission point 1. The numbers label the consecutive zeros; a ``$+$'' means that
the ray on which the zero lies had $k_o^{\varphi} > 0$, a ``$-$'' means
$k_o^{\varphi} < 0$. Projections of 3 rays on the plane of the figure are shown
in addition, to clarify the ordering of the zeros on the rays. Ray 197 is the
first one with four $\theta$ zeros,\footnote{The image of Ray 197 is
discontinued at $\varphi \approx \pi$ to avoid clogging the picture.} Ray 207 is
the first one with 6 zeros, Ray 254 is the last one with 6 zeros. The arcs of
5th zeros (drawn in $+$-s and $\times$-s) and 6th zeros partly overlap in this
projection. The arrows point to the loci of the 5th and 6th $\theta$ zeros on
Ray 207 and of the 6th $\theta$ zero on Ray 254 (Appendix \ref{graphs} explains
why the 6th zeros seem to lie beyond the endpoints of the rays.)

\begin{figure}[h]
\begin{center}
 ${}$ \\ [-5mm]
 \includegraphics[scale = 0.6]{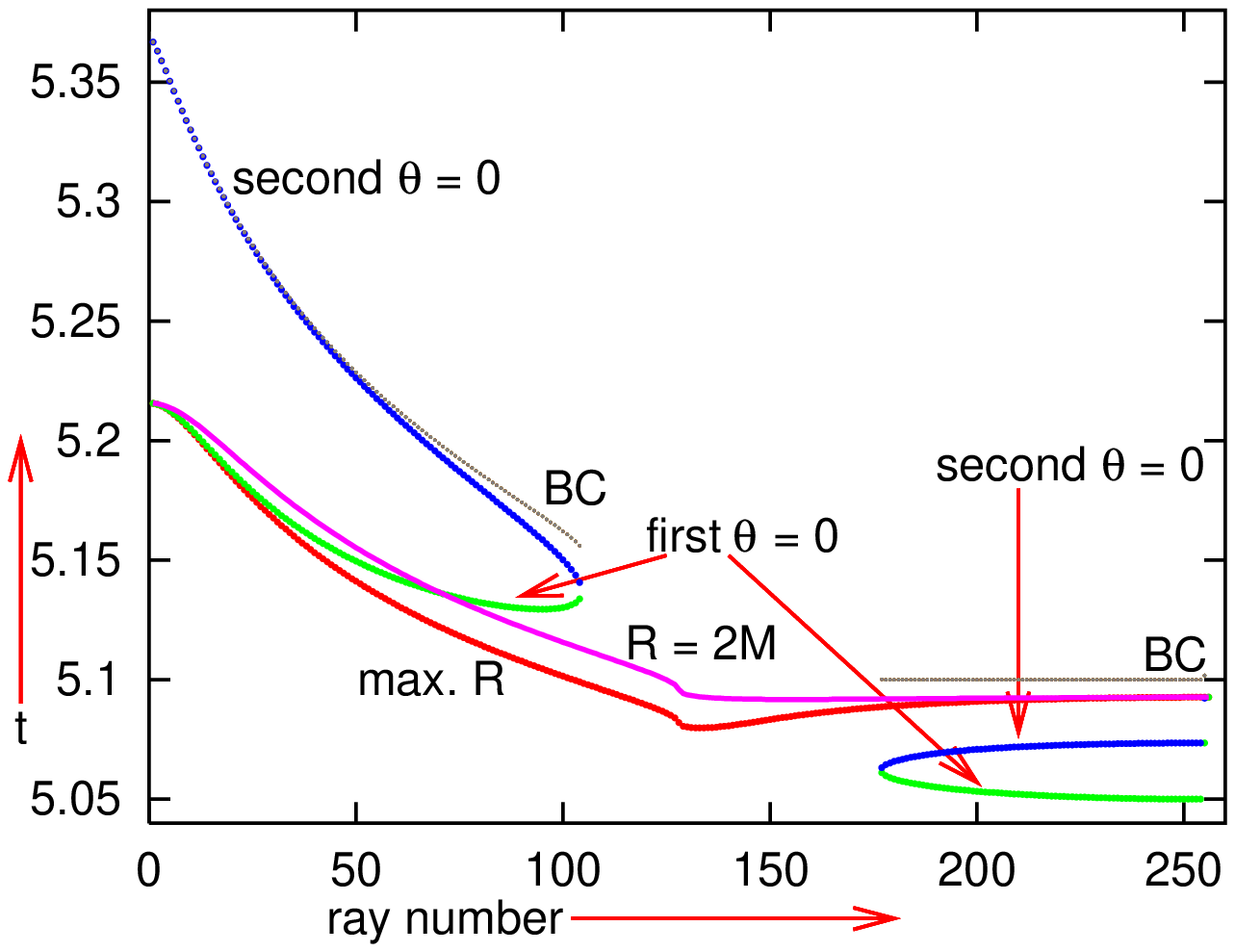}
\caption{The analogue of Fig. \ref{drawtimes} for the bundle of rays going off
the point $(M, t) = (0.005, 5.05)$. See comments in the text. }
 \label{drawtimesmale}
\end{center}
\end{figure}

\begin{figure}[h]
 ${}$ \\ [-5mm]
 \hspace{-5mm} \includegraphics[scale=0.55]{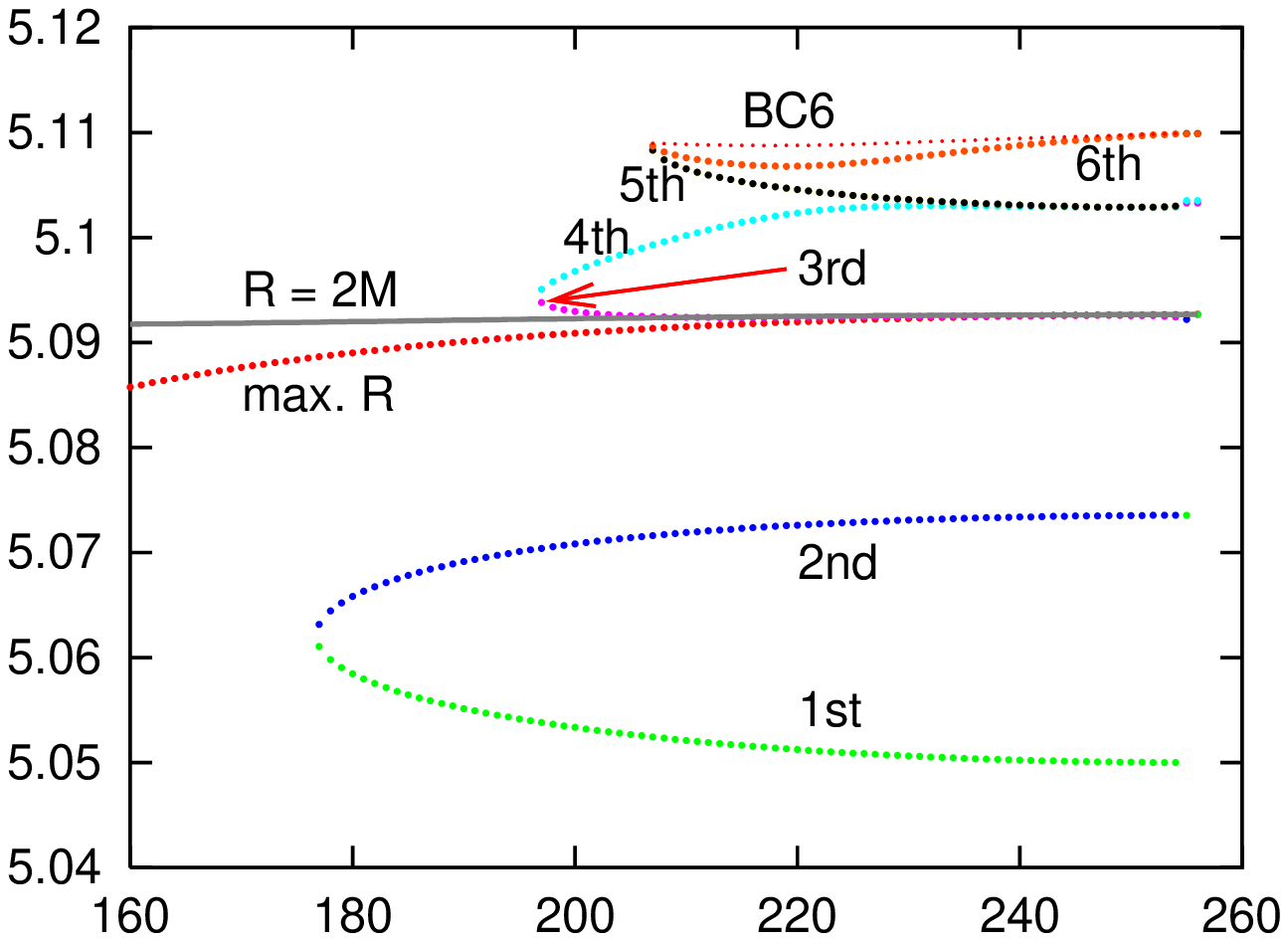}
 \hspace{-7mm} \includegraphics[scale=0.6]{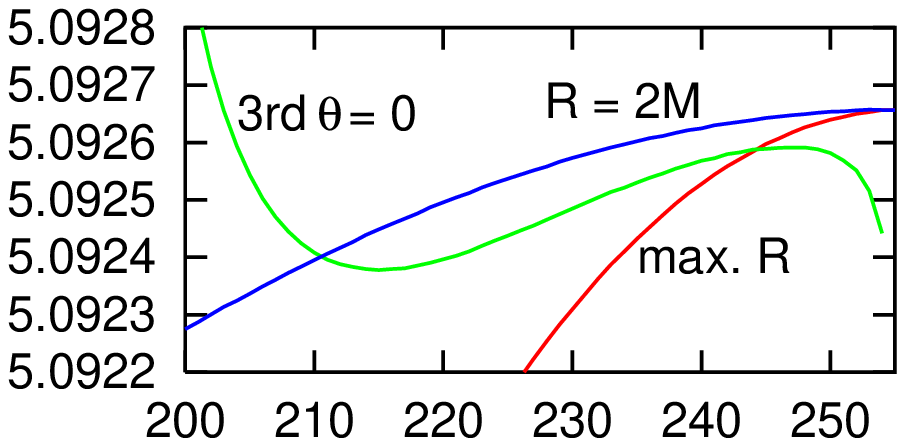}
\caption{{\bf Left:} The lower right corner of Fig. \ref{drawtimesmale}
enlarged. See comments in the text. {\bf Right:} Enlarged view on the
intersection region between $R = 2M$ and the third $\theta = 0$. }
 \label{drawtimesmalelupa}
\end{figure}

Figure \ref{drawtimesmale} is analogous to Fig. \ref{drawtimes}. The loci of the
6th $\theta$ zeros approach the BC when $\alpha_j \to \pi$. The dotted arc
marked BC6 in the left panel of Fig. \ref{drawtimesmalelupa} contains the $t$
coordinates of the BC at those $M$ where the rays pass the locus of the 6th
zero. The loci of second zeros approach the BC when $\alpha_j \to 0$. The arc of
third zeros is distinct from $R = 2M$ except at the single intersection point,
see the right panel.

\begin{figure}[h]
\begin{center}
\includegraphics[scale=0.6]{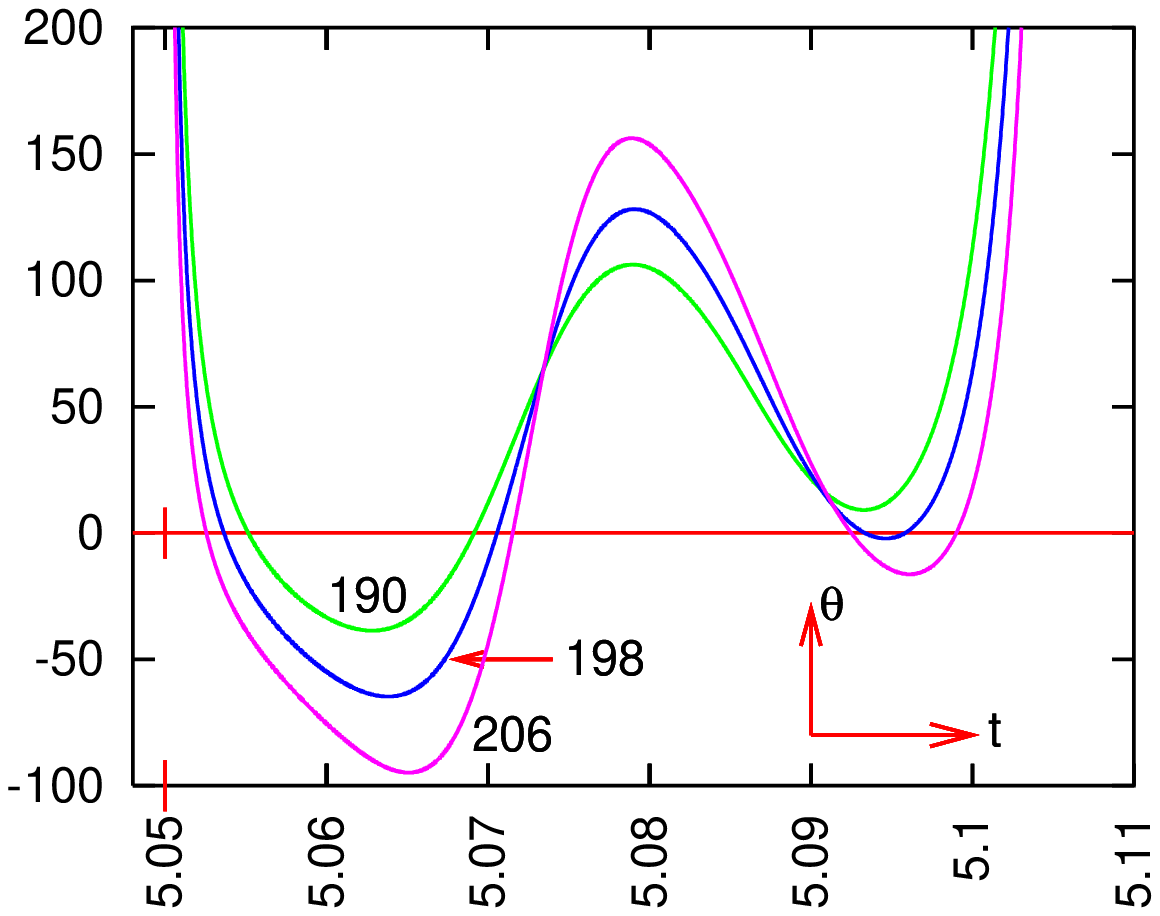}
\caption{Profiles of $\theta(t)$ along Rays 198 and 206 on which $\theta$ has 4
zeros. Ray 190, where $\theta$ has 2 zeros, is shown for comparison. See remarks
in the text.}
 \label{drawmaleth04}
\end{center}
\end{figure}

Figure \ref{drawmaleth04} shows profiles of $\theta(t)$ along two rays with 4
$\theta$ zeros. On the rays with two $\theta$ zeros, the profiles of $\theta(t)$
are similar, except that $\theta > 0$ at the second minimum, like on Ray 190.
Somewhere between Rays 197 and 198 there is one on which $\theta$ has 3 zeros.

Figure \ref{drawmaleth06} shows $\theta(t)$ profiles along exemplary rays with
six $\theta$ zeros. Between Rays 206 and 207 there is one with 5 zeros. On Ray
220, to the right of the 4th zero, $\theta$ has a jump from $31,243$ to
$-60,253$. This is interpreted as a continuous change, too rapid to be
faithfully followed by the Fortran program. The $t \approx 5.1045$ coordinate of
this jump is marked with the vertical stroke; it is the locus of the 5th
$\theta$ zero on this ray.

\begin{figure}[h]
\begin{center}
 \hspace{14cm} \includegraphics[scale=0.7]{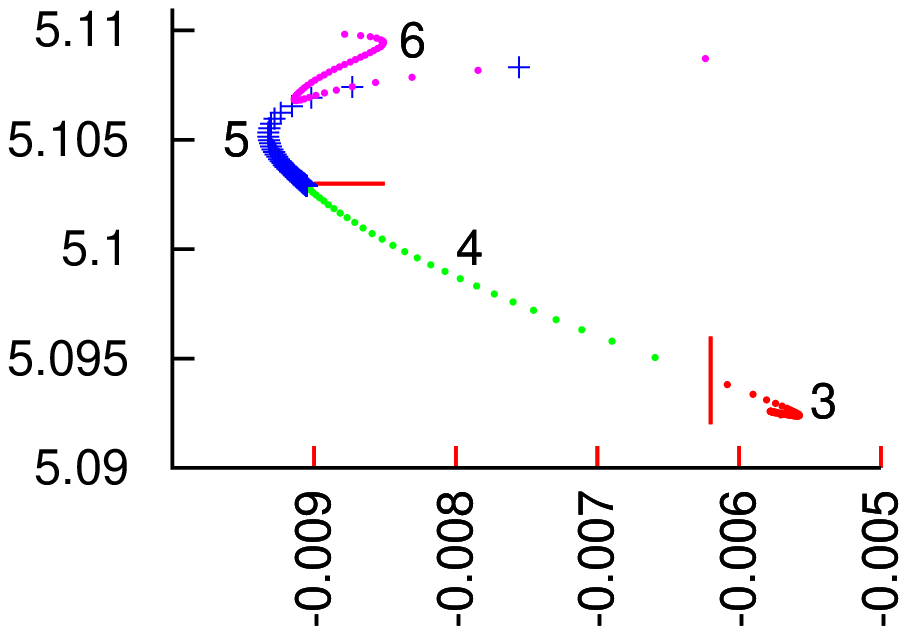}
 ${}$ \\ [-2.5cm]
 \hspace{-7.5cm} \includegraphics[scale=0.7]{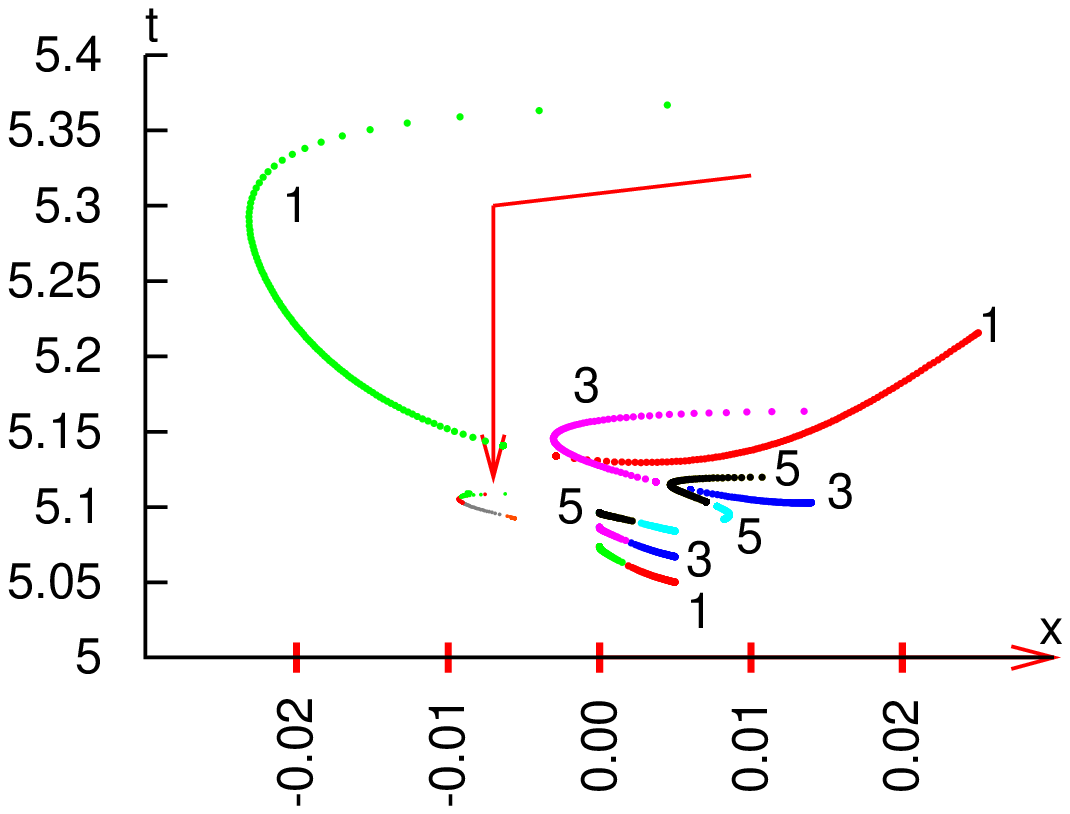}
 ${}$ \\ [-1cm]
\caption{{\bf Main panel:} The contours corresponding to emission points 1, 3
and 5 from Fig. \ref{drawzeramaleM} projected on the $y = M \sin \varphi = 0$
plane. Labels refer to the emission points. {\bf Inset:} Enlarged view on the
loci of zeros \# 3, $\dots$, 6 for the earliest emission point, in the same
projection. Here the labels refer to the consecutive number of a $\theta$-zero.}
 \label{drawth03dfront}
\end{center}
\end{figure}
\begin{figure}[h]
 ${}$ \\ [1mm]
\begin{center}
 \hspace{-4cm} \includegraphics[scale=0.7]{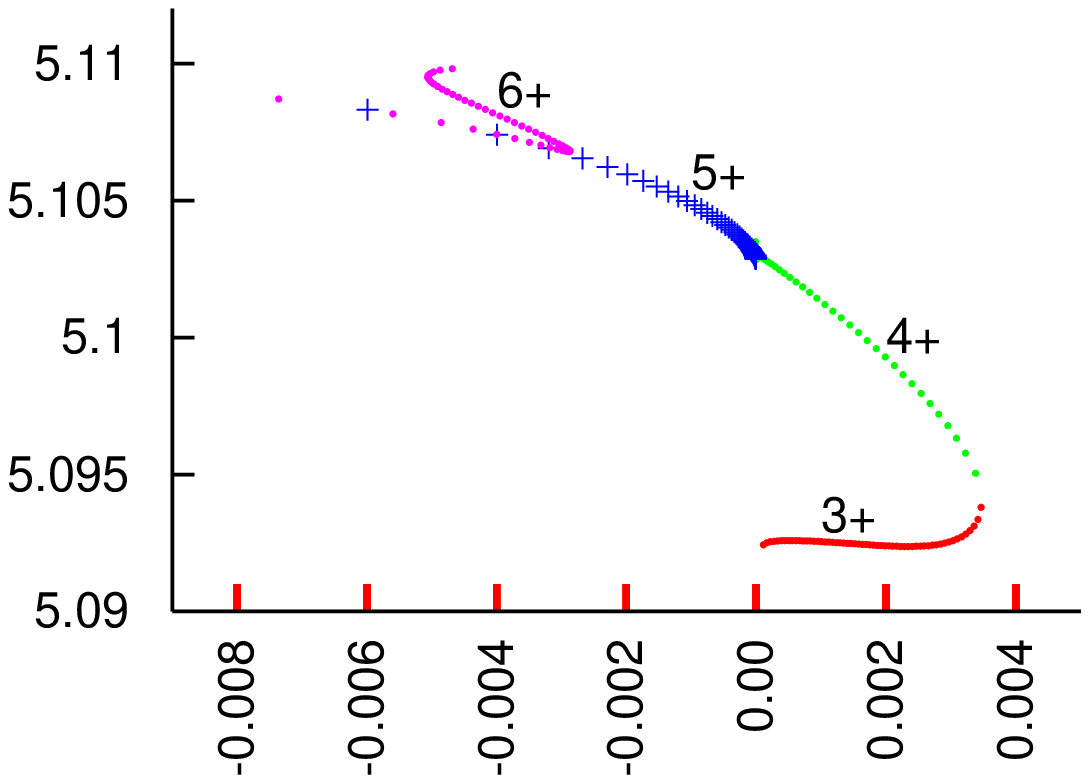}
 ${}$ \\ [-1cm]
 \hspace{-4cm}  \includegraphics[scale=0.7]{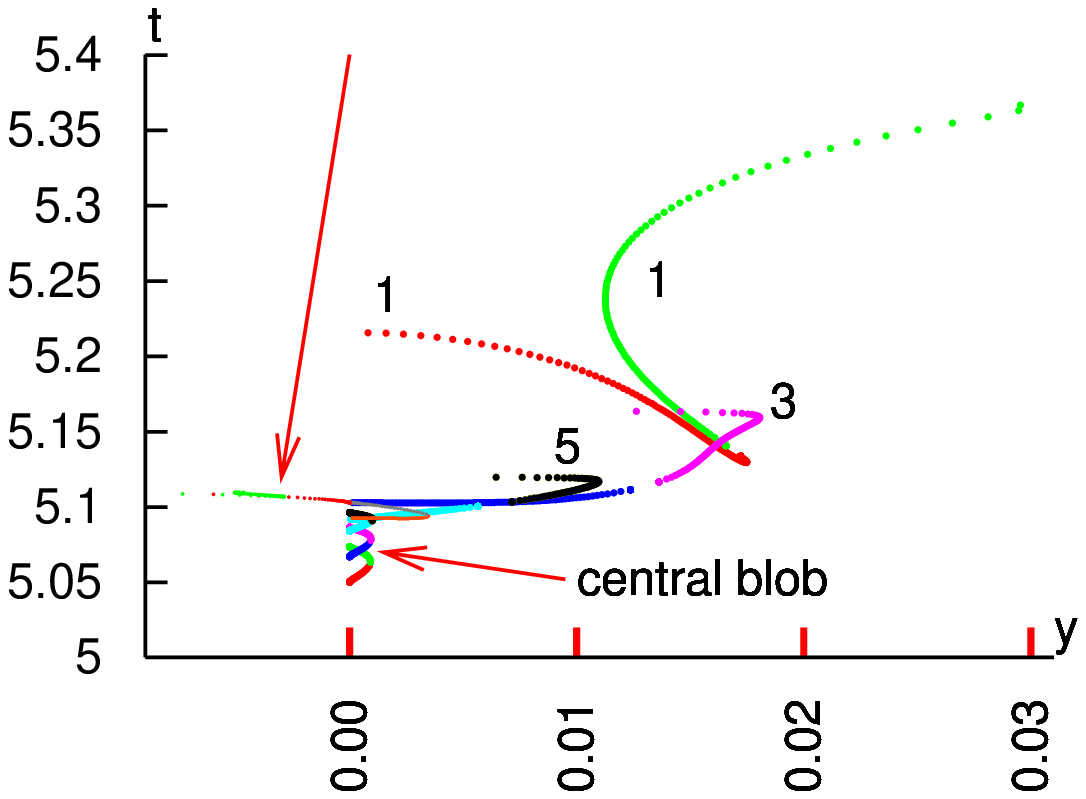}
 ${}$ \\ [-1.5cm]
\caption{The $y \geq 0$ halves of the $\theta = 0$ contours from Fig.
\ref{drawth03dfront} and the loci of zeros 3, $\dots$, 6 for emission point 1
projected on the $x = M \cos \varphi = 0$ plane. }
 \label{drawth03dside}
\end{center}
\end{figure}

Figures \ref{drawth03dfront} and \ref{drawth03dside} show selected curves from
Fig. \ref{drawzeramaleM} projected on the $y = M \sin \varphi = 0$ and $x = M
\cos \varphi = 0$ coordinate planes, respectively. They are the $\theta = 0$
contours corresponding to emission points 1, 3 and 5 (set I) and the contours of
zeros \# 3, $\dots$, 6 for emission point 1 (set II). The inset in Fig.
\ref{drawth03dfront} is a closeup view on set II. The vertical bar is at the
border between the loci of the 3rd and 4th zeros. The loci of the 4th and 5th
zeros partly overlap, the overlap zone is marked with the horizontal bar. Figure
\ref{drawth03dside} shows the $y \geq 0$ half of set I and all of set II
projected on the $x = M \cos \varphi = 0$ plane.

\section{Summary and conclusions}\label{sumcon}

\setcounter{equation}{0}

The aim of this paper was to calculate the loci of maximum $R$ and of $\theta
\equiv {k^{\mu}};_{\mu} = 0$ for bundles of rays sent from noncentral events in
L--T models ($k^{\mu}$ is the tangent vector field to the rays). It turned out
that the apparent horizon (AH) of the central observer, located at $R = 2M$,
still plays the role of the AH for noncentral observers, at least in the
exemplary model introduced in Sec. \ref{Rextreexam}. The loci of (1) maximum $R$
and of (2) $\theta = 0$ of noncentral observers do not play the role of one-way
membranes for light rays, while (3) $R = 2M$ does. This is a summary of the
reasoning that led to this conclusion:

Sections \ref{intro} -- \ref{LTnullgeo} introduced general preliminaries.

In Sec. \ref{Rextre}, the equation defining a local extremum of the areal radius
$R$ along a light ray was derived and discussed for a general L--T model. It was
shown that on nonradial rays it can exist only in the $R > 2M$ region. If it
occurs in $R < 3M$, then it is a maximum. In $R > 3M$ both minima and maxima are
possible (but may not exist).

In Sec. \ref{Rextreexam} the results of Sec. \ref{Rextre} were applied to the
exemplary toy model (ETM) introduced in Refs. \cite{KrHe2004b,PlKr2006}. The
loci of $R$ extrema were numerically calculated for rays originating at selected
events on two exemplary noncentral cosmic dust world lines in the recollapse
phase of the model and running in the equatorial hypersurface $\vartheta =
\pi/2$ (EHS). On some nonradial rays $R$ simply decreases to 0 achieved at the
BC with no extrema. On some other rays, $R$ has only maxima, on still other ones
it has both minima and maxima. The latter can happen when the ray leaves the
light source toward decreasing $R$ (which is impossible when the source is at
the center where $R = 0$).

In Sec. \ref{extexp}, the equation of the locus of $\theta = 0$ for a bundle of
light rays in a general L--T model was derived. Except on outward radial rays,
this locus is different from that of an extremum of $R$. To calculate $\theta$
along a nonradial ray numerically, an auxiliary nearby ray is needed because the
derivative ${k^r},_r$ in the formula for $\theta$ goes across the bundle. On
radial rays $\theta$ is determined by quantities intrinsic to a single ray.

In Sec. \ref{exemzerotheta}, the results of Sec. \ref{extexp} were applied to
the discussion of $\theta$ along rays running in the EHS of the same ETM that
was used in Sec. \ref{Rextreexam}. The origins of the ray bundles here lie along
the same two cosmic dust world lines as those considered in Sec.
\ref{Rextreexam}. For a given initial point, $\theta$ has typically no zeros or
two zeros along a ray, and becomes $+ \infty$ at the Big Crunch (BC). The only
rays on which $\theta \to - \infty$ at the BC are the radial ones. The other
exceptional rays are those on the boundaries between the no-zeros and the
two-zeros regions: along each of them $\theta$ has one zero, but still tends to
$+ \infty$ at the BC. When the emitter is close to the center, $\theta$ has 4 or
6 zeros along rays passing by the center (resp. 3 or 5 on the boundary rays).
The locus of the last $\theta$-zero approaches the BC when the initial direction
of the ray approaches radial.

The $\theta(t)$ profile on outward radial rays starts positive, monotonically
decreases, goes through only one zero, and tends to $- \infty$ at the BC. This
signifies focussing to a point at the BC. On inward radial rays, $\theta$ starts
negative and monotonically decreases to $- \infty$ at the BC. On other rays,
$\theta$ starts positive and initially decreases, but then becomes increasing
(after going through a minimum or more extrema) and tends to $+ \infty$ on
approaching the BC. This signifies an infinite {\em divergence} of the rays near
the BC.

Temporal orderings of loci (1) -- (3) in the EHS of the ETM were determined. The
locus of $\theta = 0$ may lie earlier or later than $R = 2M$ and than the
maximum of $R$, depending on the initial direction of the ray. These orderings
have a physical meaning. At a point where $\theta = 0$ at $t = t_{\theta = 0} <
t_{\rm R = 2M}$, an outward radial ray will go some distance toward larger $R$.
Points with $t_{\rm R = 2M} \leq t < t_{\theta = 0}$ had been isolated from the
outside world before $\theta$ became zero. This shows that for noncentral
observers the locus of $R = 2M$ rather than that of $\theta = 0$ is a one-way
membrane. Since the locus of maximum $R$ has $t_{\rm maxR} < t_{\rm R = 2M}$ on
all nonradial rays, points in the segment $t_{\rm maxR} < t < t_{\rm R = 2M}$
are not yet isolated from the communication with the outside world.

In Figs. \ref{drawzeradrugieM} and \ref{drawzeramaleM}, the intersections of
trapped surfaces \cite{Seno2011} with the $\vartheta = \pi / 2$ hypersurface
would lie between the first and second zero of $\theta$. However, $\theta < 0$
only on finite segments of some rays. Thus, if the trapped surface were evolved
into the future along these rays, its intersection with $\vartheta = \pi / 2$
would become untrapped after a finite time. Along many rays $\theta > 0$ all the
way. On those rays where $\theta < 0$ for a while, it becomes positive
eventually, going to $+ \infty$ on approaching the BC. Moreover, there exist
points on some rays where $\theta < 0$ but $R > 2M$, so they are visible from
outside -- see above and Figs. \ref{drawtimes} and \ref{drawtimesmale}. All this
shows that the formation of a trapped surface is not the ultimate signature of a
black-hole-in-the-making in situations relevant to astrophysics.

So, finally, the hypersurface $R = 2M$ does have a universal meaning in a
collapsing L--T model: this is the apparent horizon for all observers that
signifies the presence of a black hole behind it. (This meaning of $R = 2M$ was
identified by Barnes \cite{Barn1970} and Szekeres \cite{Szek1975} by considering
spherical trapped surfaces surrounding the center of symmetry and the origin,
respectively.) Events in the region $R < 2M$ are cut off from communication with
the $R > 2M$ part of the spacetime. See Ref. \cite{KrHe2004b} for an example of
how an L--T model (actually, of the same family as in Sec. \ref{Rextreexam}) can
be applied to the description of a formation of a black hole inside a spherical
condensation of dust. This conclusion shows that the transition from an L--T
model to the Friedmann (F) limit is discontinuous in one more way: the
individual AHs of noncentral observers appear abruptly. (The other discontinuity
is the abrupt disappearance of blueshifts in the F limit, first pointed out by
Szekeres in another paper \cite{Szek1980}.)

\appendix

\section{Details of Fig. \ref{duzeringi}} \label{M012}

\setcounter{equation}{0}

Rays 0 and 256 are radial (outward and inward, respectively); on them $C = 0$.
Rays 1 to 127 have $C > 0$ and $k_o^r > 0$ at the initial points, Ray 128 has $C
> 0$ and $k_o^r = 0$, Rays 129 to 255 have $C > 0$ and $k_o^r < 0$. Rays 257 to
511 are mirror images of 1 -- 255.

The exact values of $k_o^r$ are determined by $C / R_o$ via (\ref{3.12}). The
maximum $|C| / R_o = 1$ is on Rays 128 and 384. On Rays 0 -- 128, $C / R_o = j /
128$ where $j = 0, 1, \dots, 128$. On Rays 128 -- 512, $C / R_o = i /128$ where
$i = 128, 127, \dots, 1, 0$. The loci of maxima of $R$ on Rays 257 -- 511 need
not be calculated separately, they are found by inverting the signs of $y = M
\sin \varphi$ of the loci found for Rays 1 -- 255.

\section{Numerical calculation of ${k^r},_r$ in (\ref{6.7})} \label{numerics}

\setcounter{equation}{0}

At the initial point, both geodesics referred to in (\ref{6.7}) begin with $r_2
= r_1$, but $k^r_2 \neq k^r_1$. Consequently, the initial ${k^{\mu}};_{\mu} =
\infty$, so the calculation is started at step 2.

The following method was used to find $p_2$ on $G_2$ with the same $t = t_1$ as
$p_1$ on $G_1$:

First, the path of the auxiliary nearby ray $G_2$ is calculated, thereby the
collection of values of $t_n, r_n$ and $k_n^r$, $n = 1, \dots, N$, along $G_2$
is found.

Given $t$ at $p_1$ on $G_1$ we find the largest $t_n \leq t$ on $G_2$, and the
corresponding $r_n$ and $k_n^r$ on $G_2$. Then we extrapolate to $t$ by
\begin{equation}\label{b.1}
r_2 = r_n + \frac {r_{n+1} - r_n} {t_{n+1} - t_n} \left(t - t_n\right),
\end{equation}
and similarly for $k^r$. The $r_2$ and $k_2^r$ are then used in (\ref{6.7}).

\section{The sign of an infinite jump of ${k^r},_r$ and $\theta$} \label{jump}

\setcounter{equation}{0}

This is the proof that an infinite jump of ${k^r},_r$ and $\theta$ on a light
ray can only be from $- \infty$ to $+ \infty$. We assume that $- \infty < k^r <
+ \infty$ all along the ray. In (\ref{6.6}) ${k^r},_r$ is multiplied by a
non-negative coefficient, so the sign of an infinite jump in $\theta$ must be
the same as that in ${k^r},_r$.

Suppose that ${k^r},_r > 0$ at a point where $r_2 \approx r_1$. Then either $r_2
> r_1$ and $k^r_2 > k^r_1$ or $r_2 < r_1$ and $k^r_2 < k^r_1$. In the first
case, $r_2$ increases faster along the ray than $r_1$, so, as long as both
inequalities hold, $r_2 - r_1 \neq 0$. In the second case $r_2$ decreases faster
than $r_1$, with the same conclusion. In both cases $\left|{k^r},_r\right| <
\infty$.

Now suppose that ${k^r},_r < 0$ at $r_2 \approx r_1$. Then either $r_2 > r_1$
and $k^r_2 < k^r_1$ or $r_2 < r_1$ and $k^r_2 > k^r_1$. In the first case, $r_1$
increases faster along the ray than $r_2$, so may catch up with $r_2$. In the
second case the roles of $r_1$ and $r_2$ are reversed and the same conclusion
follows. In both cases ${k^r},_r$ will jump from $- \infty$ to $+ \infty$.
$\square$

\section{A comment on Fig. \ref{drawzeramaleMknot}} \label{graphs}

\setcounter{equation}{0}

In Fig. \ref{drawzeramaleMknot}, on Rays 207 and 254 the last dot marking the
6th zero of $\theta$ lies beyond the end of the ray projection. Here is the
reason of the spurious paradox: for the ray paths, one in 100 calculated data
points is shown in the figure. This is because the program drawing the graphs
could not handle the large numbers of data points that were actually calculated
(several $\times 10^5$ in some cases). So, these last points indeed do lie on
the calculated ray paths, only the paths were not interpolated to them in the
figure.

\bigskip

{\bf Acknowledgements} The idea of this paper was born in a discussion (by
email) with Jos\'e Senovilla, in connection with Ref. \cite{BeJS2013}. For some
calculations, the computer algebra system Ortocartan \cite{Kras2001,KrPe2000}
was used.


\begin{thebibliography}{99}
\bibitem{HaEl1973} S. W. Hawking and G. F. R. Ellis: {\it The
Large-scale Structure of Spaceetime}. Cambridge University Press, Cambridge
1973.

\bibitem{Lema1933} G. Lema\^{\i}tre: L'Univers en expansion [The expanding
Universe]. {\it Ann. Soc. Sci. Bruxelles} {\bf A53}, 51 (1933); English
translation: {\it Gen. Relativ. Gravit.} {\bf 29}, 641 (1997); with an editorial
note by A. Krasi\'nski: {\it Gen. Relativ. Gravit.} {\bf 29}, 637 (1997).

\bibitem{Tolm1934} R. C. Tolman: Effect of inhomogeneity on cosmological
models. {\it Proc. Nat. Acad. Sci. USA} {\bf 20}, 169 (1934); reprinted: {\it
Gen. Relativ. Gravit.} {\bf 29}, 935 (1997); with an editorial note by A.
Krasi\'nski, in: {\it Gen. Relativ. Gravit.} {\bf 29}, 931 (1997).

\bibitem{KrHe2004b}A. Krasi\'nski and C. Hellaby: Formation of a galaxy with a
central black hole in the Lemaitre -- Tolman model. {\it Phys. Rev.} {\bf D69},
043502 (2004).

\bibitem{PlKr2006} J. Pleba\'nski and A. Krasi\'nski: {\it An Introduction to
General Relativity and Cosmology}. Cambridge University Press 2006, 534 pp.

\bibitem{Frie1922} A. A. Friedmann: \"{U}ber die Kr\"{u}mmung des Raumes [On
the curvature of space], {\it Z. Physik} {\bf 10}, 377 (1922); \"{U}ber die
M\"{o}glichkeit einer Welt mit konstanter negativer Kr\"{u}mmung des Raumes [On
the possibility of a world with constant negative curvature of space], {\it Z.
Physik} {\bf 21}, 326 (1924). English translation of both papers {\it Gen.
Relativ. Gravit.} {\bf 31}, 1991 and 2001 (1999), with an editorial note by A.
Krasi\'nski and G.F.R. Ellis, {\it Gen. Relativ. Gravit.} {\bf 31}, 1985 (1999);
addendum: {\it Gen. Relativ. Gravit.} {\bf 32}, 1937 (2000).

\bibitem{Perl2004} V. Perlick: Gravitational lensing from a spacetime
perspective. {\it Living Rev Relativ.} {\bf 7} (1): 9 (2004).

\bibitem {HeLa1985} C. Hellaby and K. Lake: Shell crossings and the Tolman
model. {\it Astrophys. J.} {\bf 290}, 381 (1985) [+ erratum: {\it Astrophys. J.}
{\bf 300}, 461 (1985)].

\bibitem{Seno2011}  J. M. M. Senovilla: Trapped surfaces. {\it Int. J. Mod.
Phys.} {\bf D20}, 2139 (2011).

\bibitem{Barn1970} A. Barnes: On gravitational collapse against a cosmological
background, {\it J. Phys.} {\bf A3}, 653 (1970).

\bibitem{Szek1975} P. Szekeres: Quasispherical gravitational collapse, {\it
Phys. Rev.} {\bf D12}, 2941 (1975).

\bibitem{Szek1980} P. Szekeres: Naked singularities, in: {\it Gravitational
Radiation, Collapsed Objects and Exact Solutions}. Edited by C. Edwards.
Springer (Lecture Notes in Physics, vol. 124), New York, pp. 477 -- 487 (1980).

\bibitem{BeJS2013} I. Bengtsson, E. Jakobsson and J. M. M. Senovilla: Trapped
surfaces in Oppenheimer -- Snyder black holes. {\it Phys. Rev.} {\bf D88},
064012 (2013).

\bibitem{Kras2001} A. Krasi\'nski: The newest release of the Ortocartan set of
programs for algebraic calculations in relativity. {\it Gen. Relativ. Gravit.}
{\bf 33}, 145 (2001).

\bibitem{KrPe2000} A. Krasi\'nski, M. Perkowski: {\it The system ORTOCARTAN --
user's manual}. Fifth edition, Warsaw 2000.
\end{thebibliography}
\end{document}